\documentclass[aps, twocolumn,nofootinbib,
preprintnumbers,superscriptaddress]{revtex4-1}

\usepackage{amsmath,amssymb,amsfonts,amsthm,graphicx, epsf,dsfont,dcolumn,yfonts}
\usepackage{latexsym}
\usepackage[colorlinks=true, pdfstartview=FitV, linkcolor=black, citecolor=black, urlcolor=black]{hyperref}

\usepackage{color}

\usepackage{slashed} 

\newcommand{\be}{\begin{equation}}
\newcommand{\ee}{\end{equation}}
\newcommand{\bea}{\begin{eqnarray}}
\newcommand{\eea}{\end{eqnarray}}
\def\ie{{\it i.e.~}}

\newcommand{\bwt}{\begin{widetext}}
\newcommand{\ewt}{\end{widetext}}
\newcommand{\bo}{\raise-1mm\hbox{\Large$\Box$}}

\def\b{\beta}

\def\d{\delta}
\def\e{\epsilon}

\begin{document}

\title{Neutral Order Parameters in Metallic Criticality in {\emph d}=2+1\\
 from a Hairy Electron Star}
\author{Mohammad Edalati, Ka Wai Lo and  Philip W. Phillips}
\affiliation{ Department of Physics,
University of Illinois at Urbana-Champaign, Urbana IL 61801, USA}
\vspace{15pt}

\begin{abstract}

We use holography to study the spontaneous condensation of a neutral order parameter in a (2+1)-dimensional field theory at zero-temperature and finite density,  dual to the electron star background of Hartnoll and Tavanfar. An appealing feature of this field theory is the emergence of an IR Lifshitz fixed-point with a finite dynamical critical exponent $z$, which is due to the strong interaction between critical bosonic degrees of freedom and a finite density of fermions (metallic quantum criticality). We show that under some circumstances the electron star background develops a neutral scalar hair whose holographic interpretation is that the boundary field theory undergoes a quantum phase transition, with a Berezinski-Kosterlitz-Thouless character, to a phase with a neutral order parameter. Including the backreaction of the bulk neutral scalar on the background, we argue that the two phases across the quantum critical point have different $z$, a novelty that exists in certain quantum phase transitions in condensed matter systems. We also analyze the system at finite temperature and find that the phase transition becomes, as expected,  second-order. Embedding the neutral scalar into a higher form, a variety of interesting phases could potentially be realized for the boundary field theory. Examples which are of particular interest to condensed matter physics include an antiferromagnetic phase where a vector condenses and break the spin symmetry, a quadrupole nematic phase which involves the condensation of a symmetric traceless tensor breaking rotational symmetry, or different phases of a system with competing order parameters. 
\end{abstract}

\maketitle

\section{Introduction}

Condensed matter physics is replete with examples of ordered states
described by a neutral operator acquiring a non-zero vacuum expectation
value.  Systems exhibiting magnetism \cite{hf3},
itinerant or otherwise, or more complicated ordered phases such as quadrupolar
nematic states \cite{nematic} 
are just a few examples of such ordered states. Our focus in this paper is to use holography to describe the onset of such neutral ordered states in itinerant-like
$d=2+1$ dimensional strongly-correlated systems with a finite dynamical exponent $z$.  The standard method for treating the onset of
magnetism in a Fermi liquid is to introduce a neutral bosonic
order parameter at the level of a Hubbard-Stratonovich field in some
fermionic model and then integrate out the
fermions \cite{hertz,millis}. Because the fermions are gapless, and
hence inherently part of the low-energy degrees of freedom, they should not {\it a
  priori} be integrated out in the Wilsonian sense.  Indeed, it is now
well known \cite{ac,ssl,ms} that this procedure is highly problematic
in two spatial dimensions.  For example, in the case of $d=2+1$ and $z=2$, the fermions introduce an
infinite number \cite{ac} of non-local marginal perturbations rendering any
truncation of the resultant bosonic action at the Gaussian level moot.  In the
renormalization group sense, the critical theory in terms of the neutral bosonic order parameter does not possess any coupling
constant from which a controlled expansion can be obtained. It turns
out that even a $1/N$ expansion is of no use \cite{ssl,ms} as the naive
power counting fails, and an infinite number of planar diagrams
contributes to the fermionic self energy.  Recent
attempts \cite{Iqbal:2010eh} have been made to shed light on this problem by using holography where a strongly coupled quantum theory is
mapped onto a weakly interacting dual theory of gravity\footnote{See \cite{Hartnoll:2009sz,Herzog:2009xv,McGreevy:2009xe,Horowitz:2010gk} for reviews of holography with applications to strongly-correlated condensed matter-like phenomena.}. The boundary theory considered in \cite{Iqbal:2010eh} is dual to the Reissner-Nordstr\"om AdS$_{d+1}$  black hole  background (RN-AdS$_{d+1}$) and flows in the IR to a (0+1)-dimensional CFT with a dynamical exponent $z=\infty$. Hence, such a theory may have less direct bearing on real condensed matter systems, compared to a theory which flows in the IR to a fixed-point with finite $z$.  Another drawback of this theory is the existence of a finite entropy density at zero temperature.

All is not lost with holography in this context, however. Indeed,
holography is ideally suited to treating this problem as the failures of the existing conventional treatments \cite{ac,ssl,ms} stem from the inherent strongly-coupled nature of the underlying theory.  We show here that it is possible to construct a
holographic setup in which a neutral bosonic operator condenses in an
itinerant-like electron fluid with a finite value of $z$, and in $d=2+1$. Interestingly, the case we consider also includes $z=2$.  A finite value of $z$ in the IR is achieved by replacing the ${\rm AdS}_2\times \mathbb{R}^{d-1}$ near-horizon
geometry of the RN-AdS$_{d+1}$ background with one that is of, say, Lifshitz type \cite{Kachru:2008yh} as in \cite{Hartnoll:2010gu}.  Our work here then represents a holographic description of quantum
criticality in a metallic system in two spatial dimensions with $z=2$, and with a neutral bosonic order parameter.

It was realized in \cite{Hartnoll:2008kx} that a neutral scalar field with a mass square satisfying the Breitenlohner-Freedman (BF) bound \cite{Breitenlohner:1982bm}  of the asymptotic AdS$_{d+1}$ can cause an instability in the extremal, or near-extremal,  RN-AdS$_{d+1}$ background if its effective mass square in the AdS$_2$ region of the near horizon geometry violates the BF bound of AdS$_2$.  As a result, the RN-AdS$_{d+1}$ background could become unstable to forming a neutral scalar hair\footnote{This mechanism is identical to the one used for the construction of holographic superconductors \cite{Gubser:2008px, Hartnoll:2008vx, Hartnoll:2008kx, Denef:2009tp}. As discussed in \cite{Faulkner:2010fh, Faulkner:2010gj},  there is another way of making the AdS$_2$ region unstable even if the effective mass square of the scalar field, either charged or neutral,  satisfies the BF bound of AdS$_2$. This mechanism involves deforming the boundary theory by relevant multi-trace operators \cite{Witten:2001ua,Berkooz:2002ug}, which are, in this context, constructed out of the operator dual to the scalar field.}.  The holographic interpretation of this hairy background is that the boundary theory is in a phase where the operator dual to the bulk neutral scalar field condenses.   Such a condensate was explicitly constructed in \cite{Iqbal:2010eh}  for the boundary theory dual to the RN-AdS$_{d+1}$ background, where it was shown that the operator condenses below some critical temperature as long as its conformal dimension is in a certain range. The phase transition at finite temperature was shown to be second-order with  mean-field exponents. Moreover, in the standard quantization, the critical temperature goes to zero as the conformal dimension of the  operator approaches a critical dimension from below. At zero temperature, on the other hand, by varying the conformal dimension of  the operator, the boundary theory undergoes a quantum phase transition of the Berezinski-Kosterlitz-Thouless (BKT) type near the quantum critical point. The authors of \cite{Iqbal:2010eh} then used the condensation of the neutral scalar operator as a starting point  to construct an antiferromagnetic phase in the boundary theory. This was accomplished by embedding the bulk neutral scalar field into a triplet whose dual operator condenses in such a way as to break the SU(2) spin symmetry to U(1) (see also \cite{Faulkner:2010tq}). The fluctuations of this  antiferromagnetic order parameter (two Goldstone modes) were then analyzed in \cite{Iqbal:2010eh} and their dispersion relations were determined. 

As mentioned above, the boundary theory considered in
\cite{Iqbal:2010eh} flows in the IR to a (0+1)-dimensional CFT with a
dynamical exponent $z=\infty$.  This is an unwelcome  feature in
modeling metallic quantum criticality using holography.  In this
paper, we ask ourselves whether it is possible to condense a neutral
scalar operator  in a boundary field theory which flows in the IR to a
fixed-point with finite $z$.  We find that the answer is yes.  To
achieve this goal, one first replaces the ${\rm AdS}_2\times \mathbb{R}^{d-1}$ near horizon geometry of the RN-AdS$_{d+1}$ background with, say, a Lifshitz geometry.  Consequently, the background is asymptotically 
AdS$_{d+1}$ while in the IR it has Lifshitz scaling. For simplicity,
and with an eye towards  condensed matter applications, we assume that
the boundary theory is $d=2+1$ dimensional. A background with such
properties was recently constructed by Hartnoll and Tavanfar in \cite{Hartnoll:2010gu} following \cite{Hartnoll:2009ns, deBoer:2009wk}. As in \cite{Hartnoll:2010gu}, we refer to this background as the electron star background, or simply the star background.  Compared to other backgrounds which are asymptotically AdS$_4$ and Lifshitz in the IR \cite{Goldstein:2009cv}, the electron star background has the appealing feature that the IR Lifshitz scaling is due to the strong interaction between critical bosonic degrees of freedom and a finite density of fermions, which is of utmost importance in metallic quantum criticality.

\begin{figure}
\centering
 \includegraphics[width=87mm]{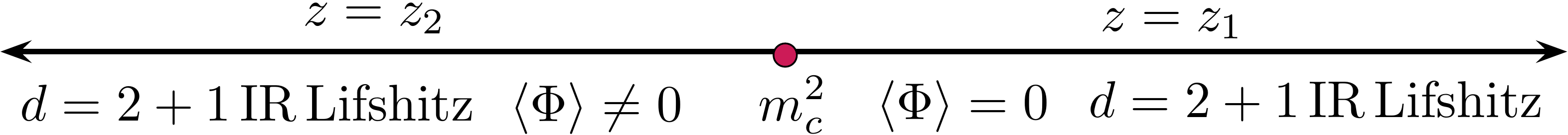}
\vskip 0.05in
\caption{\label{Phase} \footnotesize{A cartoon of the phase transition in our hologrphic setup as a function of the mass square, $m^2$, of the bulk neutral scalar field $\phi$.  Once $m^2$ violates the BF bound, $m_c^2$, of the far-interior Lifshitz region of the electron star background, the boundary theory goes into a phase where  the dual operator $\Phi$ spontaneously condenses. The backreaction of the scalar on the geometry changes the value of the dynamical exponent $z$.  The condensation of $\Phi$ is controlled by the IR characteristics of the background which remains $d=2+1$ Lifshitz across the transition.} }
\end{figure}

The (2+1)-dimensional boundary
theory dual to the electron star background is a zero-temperature, but
finite density, field theory which flows in the IR to a
(2+1)-dimensional Lifshitz fixed-point with a finite $z$. We construct
a neutral scalar hair for the electron star background and show that
upon varying the UV conformal dimension of the dual scalar
operator, or any other knob in the boundary theory  which allows the
effective mass square of the bulk scalar in the far-interior Lifshitz
region to change, the boundary field theory undergoes a quantum phase
transition to a phase with a neutral order parameter.  Figure \ref{Phase} shows a cartoon of the phase transition.  A key characteristic we find is that the dynamical exponent changes across the transition.  We analyze the nature of the phase transition at zero temperature and argue that it has a BKT character. We then consider the setup at finite temperature and show that below some critical temperature, the neutral order parameter condenses and moreover the transition is of the standard mean field type.  While the focus of this paper is mainly on the condensation of a single neutral scalar operator in the boundary theory, we discuss, very briefly, its applications for the holographic modeling of phases with vector or tensor order parameters, or systems with multiple order parameters. For example, following \cite{Iqbal:2010eh},  an antiferromagnetic phase of the boundary theory in our case can easily be constructed by embedding the bulk neutral scalar into a triplet which is charged under the SU(2) spin symmetry, and allowing the dual neutral vector operator condense and break SU(2) down to U(1). Constructing a quadrupole nematic phase, where the order parameter is a symmetric traceless tensor which breaks the SO(2) rotational symmetry, is more difficult. We comment on the challenges of constructing such a phase using the embedding of the bulk neutral scalar into a symmetric traceless  tensor. 

The paper is organized as follows. We start the next section by reviewing the electron star background of \cite{Hartnoll:2010gu}, and its finite-temperature variant \cite{Puletti:2010de, Hartnoll:2010ik}.  In section \ref{SectionThree} we consider a neutral scalar field as a probe in the electron star background and show that under some circumstances this gravitational system undergoes a phase transition to a background with a neutral scalar hair.  The holographic interpretation of the transition is that the boundary field theory undergoes a quantum phase transition of the BKT type from a $\mathbb{Z}_2$-symmetric phase to a phase with a neutral order parameter, which spontaneously breaks this symmetry. We then consider the backreaction of the neutral scalar on the star background and show that the aforementioned two phases have different dynamical exponents.  Section \ref{SectionFour}  extends the analysis to the finite temperature case.  In the last section we discuss various applications of our holographic setup and conclude with open questions and directions for future work.

\section{The Background}\label{Section Two}

The boundary theory that we wish to study is dual to the so-called  electron star background. The zero-temperature background was constructed in \cite{Hartnoll:2010gu}, and was later extended to finite temperature in \cite{Puletti:2010de, Hartnoll:2010ik}. In this section we briefly review the star background both at zero and finite temperature, and set the stage for later calculations.

\subsection{Zero-Temperature Background}

Consider an ideal fluid of charged fermions in a (3+1)-dimensional space-time with a negative cosmological constant $\Lambda=-3/L^2$.  The equations of motion of the system read 
\begin{align}\label{EinsteinEq}
R_{MN}-\frac{1}{2}R\,g_{MN}+\Lambda\,g_{MN}&=\kappa^2T_{MN},\\
\nabla^M F_{MN}&=e^2 J_N,\label{MaxwellEq}
\end{align}
where $\kappa^2=8\pi G_{\rm N}$, $e$ is the fermion charge and 
\begin{align}\label{Stress}
T_{MN}&=\frac{1}{e^2}\Big(F_{MP}F^{P}_N-\frac{1}{4}g_{MN}\,F_{PQ}F^{PQ}\Big)\nonumber\\
&+(\rho+p)\,u_Mu_N+p\,g_{MN},\\
J_M&=\sigma\,u_M,\label{Current}
\end{align}
with $\rho$, $p$, $\sigma$ and $u_a$ being, respectively, energy
density, pressure, charge density and four-velocity (satisfying the
normalization $u_M u^M =-1$) of the fluid. Throughout the paper, we denote the bulk indices
by capital letters $M,N,\dots=\{t,x,y,r\}$  while the boundary theory
directions are denoted by the Greek indices $\mu, \nu,
\ldots=\{t,x,y\}$.

To solve the above equations, we take the ansatz for the metric and the gauge field to be of the following form 
\begin{align}\label{BackgroundAnsatz}
\frac{ds^2}{L^2}&=-f(r)dt^2+\frac{1}{r^2}\left(dx^2+dy^2\right)+g(r)dr^2,\\
A&=\frac{eL}{\kappa}h(r) dt,\label{GaugeAnsatz}
\end{align}
where the asymptotic boundary is at $r\to 0$. In addition, an equation of state for the fluid must also be specified. Suppose the fluid is composed of zero temperature charged fermions with mass $m_{\rm f}$, with  the density of states   
\begin{align}\label{DensityOfStates}
n(E)=\beta E \sqrt{E^2-m_{\rm f}^2}, 
\end{align}
where $\beta$ is a constant of order one. The authors of \cite{Hartnoll:2010gu} then assumed a locally flat approximation where, at each $r$-slice, the fermion dynamics are determined by the local chemical potential 
\begin{align}\label{LocalChemicalPotential}
\mu_{\rm loc}(r)=\frac{e}{\kappa}\frac{h(r)}{\sqrt{f(r)}},
\end{align}
and argued that such an approximation is self-consistent in a regime of parameters where gravity can be treated classically with an order-one backreaction of the fermions on the geometry. Given this approximation one can then use \eqref{DensityOfStates} to compute the fluid energy density, pressure and charge density as function of $r$. Obviously, $n(E)=\rho(r)=p(r)=\sigma(r)=0$ for $r\leq r_s$ where $r_s$ is the boundary of the fluid given by solving $\mu_{\rm loc}(r_s)=m_{\rm f}$. 

It is more convenient to rescale $\rho\rightarrow \rho/(L^2\kappa^2)$, $p\to p/(L^2\kappa^2)$, $\sigma\to \sigma/(L^2e\kappa)$, $\beta\to (\kappa^2/e^4L^2)\beta$, $m_{\rm
  f}\to(e/\kappa)\,m_{\rm f}$ and $\mu_{\rm loc}\to(e/\kappa)\,\mu_{\rm loc}$, with the new quantities being all dimensionless. Putting everything together, the equations of motion then take the form
\begin{align}
\frac{1}{r}\left(\frac{f'(r)}{f(r)}+\frac{g'(r)}{g(r)}+\frac{4}{r}\right) +\frac{g(r)h(r)}{\sqrt{f(r)}}\sigma(r)&=0,\label{EintReduced}\\
\frac{f'(r)}{rf(r)}-\frac{h'(r)^2}{2f(r)}+\left[3+p(r)\right]g(r)-\frac{1}{r^2}&=0,\label{EosF}\\
h''(r)+\frac{g(r)}{\sqrt{f(r)}}\left(\frac{r}{2}h(r)h'(r)-f(r)\right)\sigma(r)&=0\label{MaxReduced}. 
\end{align}

In the far-interior region $r\to \infty$, the solution is of the Lifshitz type  \cite{Hartnoll:2010gu}. As argued there, in order for the Lifshitz solution to make sense, one should have $0\leq m_{\rm f} <1$. The solution in the interior region $r_s< r <\infty$ (where the fluid energy density, pressure and charge density are all non-zero) can easily be worked out by perturbing the Lifshitz solution, and demanding the perturbation to grow as $r\to 0$, but not blow up as $r\to\infty$. In so doing, one finds
\begin{align}
f(r)&=\frac{1}{r^{2z}}\left(1+\mathfrak{f}_1 r^{\alpha_-}+\cdots\right),\label{fPerturbed}\\
g(r)&=\frac{\mathfrak{g}}{r^{2}}\left(1+\mathfrak{g}_1 r^{\alpha_-}+\cdots\right),\label{gPerturbed}\\
h(r)&=\frac{\mathfrak{h}}{r^{z}}\left(1+\mathfrak{h}_1 r^{\alpha_-}+\cdots\right) \label{hPerturbed},
\end{align}
where $z=z(m_{\rm f},\beta)$ is the dynamical critical exponent 
\begin{align}\label{GandH}
\mathfrak{g}^2&=\frac{36z^4(z-1)}{\left[\left(1-m_{\rm f}^2\right)z-1\right]^3\beta^2}, \qquad\mathfrak{h}^2=\frac{z-1}{z},
\end{align}
and  $\mathfrak{g}_1$, $\mathfrak{h}_1$ and all higher order perturbation coefficients are determined in terms of $\mathfrak {f}_1$, which itself could be set to any value by rescaling $r$, $t$, $x$, $y$.  Also, $\alpha_-$ is determined in terms of $m_{\rm f}$ and $z$, the expression of which can be found in  \cite{Hartnoll:2010gu}. The dependence of $z$ on $m_{\rm f}$ and $\beta$ is complicated, but satisfies $z\geq (1-m_{\rm f}^2)^{-1}\geq 1$. (A numerical plot of $z$ in terms of $\beta$ for some sample values of $m_{\rm f}$, along with the explicit forms of its asymptotic expressions can also be found in \cite{Hartnoll:2010gu}.) The solution in the exterior region  $0< r \leq r_s$ (with the fluid energy density, pressure and charge density all being zero) is RN-AdS$_{4}$ where the metric and gauge field functions are given by  
\begin{align}\label{Exteriorf}
f(r)&=\frac{1}{r^2}\Big(c^2-Mr^3+\frac{1}{2}Q^2r^4\Big), \,\,\,\,\,\,\,
g(r)=\frac{c^2}{r^4 f(r)},\nonumber\\
h(r)&=\mu-Q r.
\end{align}
Matching $f(r)$, $g(r)$, $h(r)$ and $h'(r)$ of the interior and exterior solutions at $r=r_s$ will then determine the constants $c$, $M$, $Q$ and $\mu$. Note that the constant $c$ is introduced because choosing a particular value of $\mathfrak{f}_1$ in the interior solution fixes the normalization of time. The electron star background is a family of asymptotically AdS$_4$ solutions parametrized by $m_{\rm f}$ and $\beta$.  The far-interior region of the background is generically Lifshitz. But, in some limits of the $\beta$-$m_{\rm f}$ parameter space, the far-interior region is characterized by a geometry other than Lifshitz. For example, an ${\rm AdS}_2\times \mathbb{R}^2$ solution (where $z=\infty$) can be recovered either by taking $\beta \to 0$ while keeping $m_{\rm f}$ fixed, or by taking the $m_{\rm f}\to 1$ limit from below. Another interesting limit of the far-interior solution is ${\rm AdS}_4$ (with $z=1$) which is obtained by taking $\beta\to\infty$ at $m_{\rm f}=0$. 

In this paper, we exclude the two limiting ${\rm AdS}_2\times \mathbb{R}^2$ and ${\rm AdS}_4$ interior solutions by taking $m_{\rm f}\in (0,1)$. In other words, the far-interior solution is Lifshitz with a finite  $z>1$. In this case, one obtains $\mathfrak{g}>1$, which turns out to be crucial in our subsequent discussions. Note that the zero-temperature electron star background has zero entropy density.

\subsection{Finite-Temperature Background }

The details of the finite-temperature version of the electron star
background are as follows \cite{Puletti:2010de,
  Hartnoll:2010ik}. Temperature is introduced by  making the Euclidean
time coordinate periodic with a period $1/T$, where $T$ is the
temperature.  This then requires a horizon at some finite radius
$r=r_0$ in the interior of the geometry. As argued in
\cite{Puletti:2010de, Hartnoll:2010ik}, in the regime of validity of
the electron star background $e\sim \kappa/L \ll 1$, the effect of
temperature on the fermion density of state could be neglected.  Therefore, the density of states is given by \eqref{DensityOfStates} and with the same planar  ansatz for the metric and the gauge field as before, one has to solve the equations \eqref{EintReduced}--\eqref{MaxReduced} in order to determine the finite temperature background. Again, for large enough $m_{\rm f}$, the fluid energy density, pressure and charge density vanish everywhere. As a result, a RN-AdS$_4$ background will satisfy those equations. On the other hand, if $m_{\rm f}$ is small enough, there exists a critical radius $r=r_c$ for which $m_{\rm f}$ equals the local chemical potential $\mu_{\rm loc}$ at that radius. This happens at a critical temperature $T=T_*$. The quantities $r_c$ and $T_*$ are determined by solving the two equations $\mu_{\rm loc}(r_s)=m_{\rm f}$ and $\mu'_{\rm loc}(r_s)=0$, with prime denoting derivative with respect to $r$.

For $T<T_*$, there exists a finite width $r_2> r>r_1$ for which the fluid has non-zero energy density, pressure and charge density. The radii $r_1$ and $r_2$ are solutions to $\mu_{\rm loc}(r_{1,2})=m_{\rm f}$. Thus, the finite-temperature electron star background  is divided into three regions. In the so-called inner region of the background, $r_0\geq r\geq r_2$,  the solution is a RN-AdS$_4$ black hole with
\begin{align}\label{Inner Region}
f(r)&=\frac{1}{r^2}\Big(1-M_0r^3+\frac{1}{2}Q_0^2r^4\Big), \,\,\,\,\,\,\,\,g(r)=\frac{1}{r^4f(r)},\nonumber\\
h(r)&=\mu_0-Q_0r,
\end{align} 
where $r_0$ is the horizon radius given by the largest real root of $f(r)=0$,  $\mu_0=Q_0r_0$ and $M_0r_0^3=1+\mu_0^2r_0^2/2$. The temperature of the black hole is given by $T=(4\pi c)^{-1}\left| df(r_0)/dr\right|$, where the parameter $c$ is introduced for later convenience. The quantity $\mu_0$ is the inner region chemical potential which is proportional to the charge $Q_0$ hidden $``$inside$"$ the black hole horizon. Note that $\mu_0$  does not correspond to the chemical potential of the boundary field theory. Indeed, $\mu_0 r_0$ can be used as a tuning parameter to change the temperature and the chemical potential of the boundary field theory. In the intermediate region $r_2>r>r_1$, the fluid pressure,  energy density and charge density are all non-zero. The background in this intermediate region is obtained by solving the equations \eqref{EintReduced}--\eqref{MaxReduced}. In the exterior region, namely $r_1\geq r$, the background is again described by a RN-AdS$_4$ solution with 
\begin{align}
f(r)&=\frac{c^2}{r^2}\Big(1-Mr^3+\frac{1}{2}Q^2r^4\Big),\,\,\,\,\,\,\,
g(r)=\frac{c^2}{r^4f(r)},\nonumber\\
h(r)&=c\left(\mu-Qr\right),
\end{align} 
where the boundary theory quantities $c, M,Q$ and $\mu$ are determined by matching the solutions at $r=r_1$ and $r_2$. Note that $\mu$ is the chemical potential of the boundary field theory. 

In summary, for $m_{\rm f}\in(0,1)$, as the temperature is lowered, the RN-AdS$_4$ background undergoes a phase transition to a finite-temperature electron star background. It was shown in \cite{Puletti:2010de, Hartnoll:2010ik} that this phase transition is third order. Also, it is worth emphasizing that an appealing feature of the finite-temperature electron star background is that, unlike the RN-AdS$_4$ solution, the entropy density $s$ approaches zero as $s\sim T^{2/z}$ in the $T\to 0$ limit \cite{Hartnoll:2010ik}, where, for given allowed values of the parameters $m_{\rm f}$ and $\beta$, the dynamical critical exponent $z$ agrees with the one obtained from the zero-temperature electron star background with the same values of $m_{\rm f}$ and $\beta$.  

\section{Neutral Scalar Order Parameter at Zero Temperature}\label{SectionThree}

In this section we analyze the spontaneous condensation of a neutral scalar operator in a strongly-coupled (2+1)-dimensional zero-temperature boundary field theory with a finite density of fermions, which is   dual to the the zero-temperature electron star background reviewed in the previous section. In the context of the AdS/CFT correspondence, the question of whether a scalar operator can spontaneously condense in a strongly-coupled boundary theory maps to the question of whether the gravitational background can develop a scalar hair. So, we are naturally led to investigate whether the electron star background can develop a neutral scalar hair.  
 
\subsection{Neutral Scalar Hair for the Electron Star}

We introduce the scalar field $\phi$ in the electron star background
by considering the action
\begin{align}\label{Scalar Action}
S_\phi=-\frac{1}{2\kappa^2\lambda}\int d^4x \sqrt{-g}\left[\frac{1}{2}(\nabla \phi)^2+V(\phi)\right],
\end{align}
where the potential $V(\phi)$ takes the form 
\begin{align}\label{Potential}
V(\phi)=\frac{1}{4L^2}\left[\left(\phi^2+m^2L^2\right)^2-m^4L^4\right].
\end{align}
We choose the mass square of the scalar, $m^2$, to be negative but
above the BF bound of the asymptotic AdS$_{4}$ region. Namely, the mass of the scalar is assumed to be in the range $-9/4<m^2L^2<0$. Note that since $m^2$ is taken to be negative, the field $\phi$ should be expanded asymptotically around the maximum\footnote{Expanding $\phi$ in the asymptotic region around either of the two minima of $V(\phi)$ corresponds to a positive $m^2$. We will not consider such a case in this paper, as, with the mechanism explained in the paper, we do not expect to see a condensation of the dual neutral scalar operator in the boundary theory.}  of the Mexican-hat  potential $V(\phi)$, \ie $\phi=0$.  As we will explain below,  it is important that $V(\phi)$ has (at least) a minimum \cite{Iqbal:2010eh}.  Note that the constant term  in $V(\phi)$ is there to ensure zero contribution to the cosmological constant of the electron star background, $\Lambda=-3/L^2$, for $\phi=0$. Also, $\lambda$ is a coupling constant chosen, at first, to be large in order for the probe approximation to be valid. We later relax this assumption when we consider the backreaction of the scalar on the electron star background. The same scalar action  was used  in \cite{Iqbal:2010eh} to study the condensation of a neutral scalar order parameter in the boundary
theory dual to the RN-AdS$_{4}$ background, both at zero and finite
temperature\footnote{See also \cite{Faulkner:2010gj} for a discussion of condensing neutral scalar operators in holographic setups using multi-trace deformations.}.

The action \eqref{Scalar Action} is invariant under a $\mathbb{Z}_2$
symmetry: $\phi\to-\phi$. This symmetry is broken in a phase of
the boundary theory where the operator dual to $\phi$
spontaneously condenses. The symmetry group can easily be enlarged by
considering more scalar fields, and depending on which dual operators
condense, a variety of interesting symmetry breaking patterns could be
realized in the boundary theory. Such holographic setups can potentially be used to capture the relevant physics of the strongly-correlated systems with competing order parameters. In this section, we deal with just one neutral scalar field with an action given in \eqref{Scalar Action}. 

The equation of motion for $\phi$ reads
\begin{equation}\label{PhiEOM}
\frac{1}{\sqrt{-g}}\partial_M \left(g^{MN}\sqrt{-g}\,\partial_N \phi\right)=V'(\phi),
\end{equation}
where prime here denotes derivative with respect to $\phi$. Given the background metric in \eqref{BackgroundAnsatz} and choosing the ansatz $\phi=\phi(r)$, equation \eqref{PhiEOM} takes the form
\begin{align}\label{PhiEOMRefined}
\hskip-0.1in\frac{r^2}{\sqrt{f(r)g(r)}}\,\partial_r \left(\frac{1}{r^2}\sqrt{\frac{f(r)}{g(r)}}\partial_r \phi\right)=L^2V'(\phi).
\end{align}
\begin{figure*}
\centering
\hskip -0.09in\includegraphics[width=70mm]{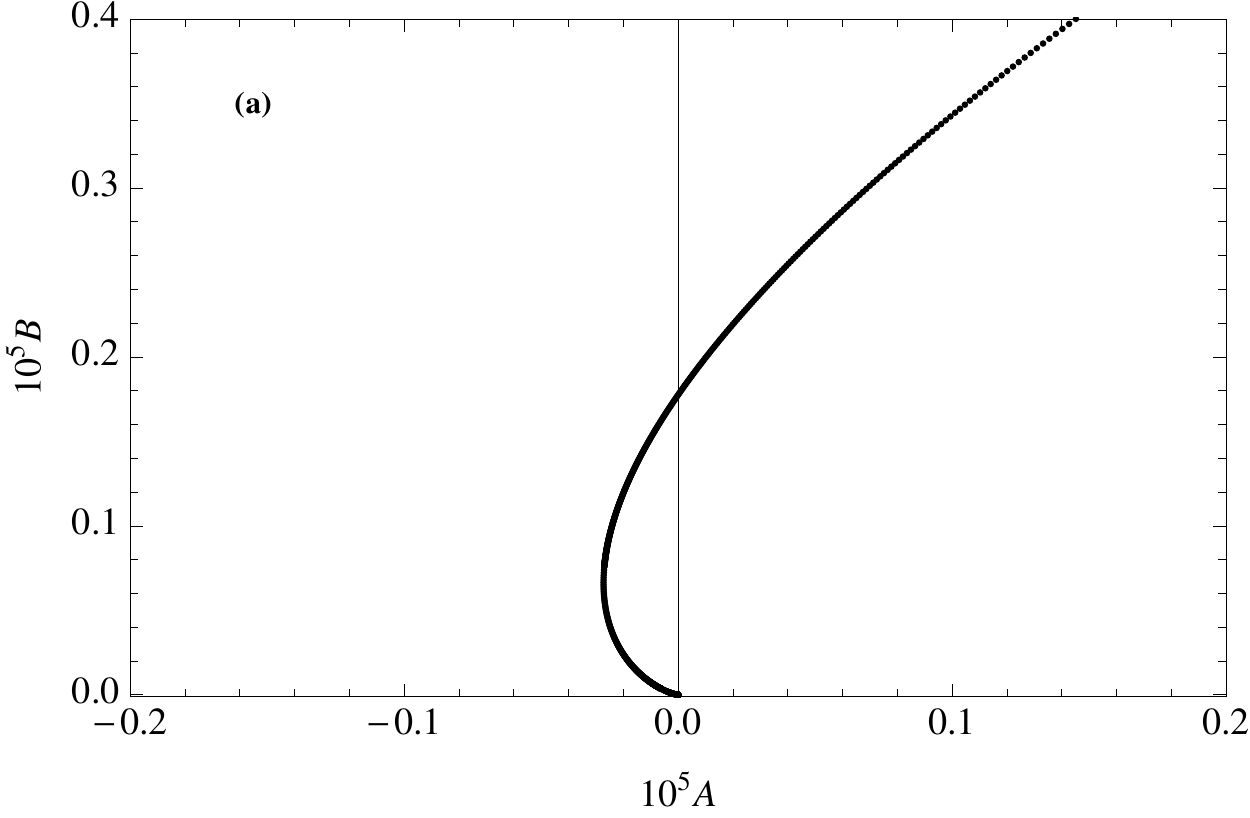}\qquad\qquad \qquad
\includegraphics[width=68mm]{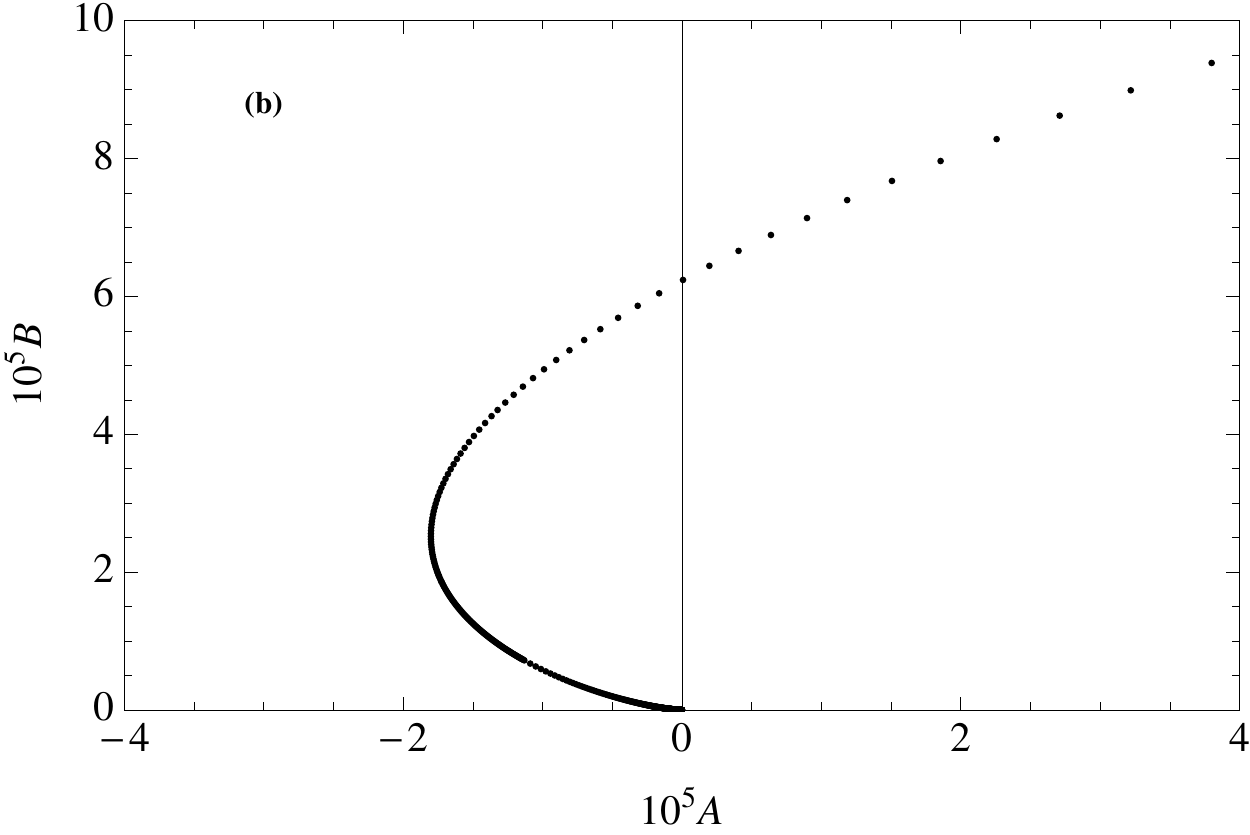}
\caption{\label{CondensateFigure} \footnotesize{Plots of $B$ versus $A$ for the neutral scalar field $\phi$ with a mass square $m^2L^2=-2.20$ on an electron star background with $m_{\rm f} =0.36$ and (a) $z=2$ and (b) $z=3$.  The BF bounds in the far-interior of the star (with $m_{\rm f}=0.36$) for $z=2$ and $z=3$ are $m_c^2L^2\simeq -2.12$ and $-2.02$, respectively. Since $B$ is non-zero at $A=0$, the plots show the spontaneous condensation of the dual operator $\Phi$ in the respective boundary theories. } }
\end{figure*}

We are after a non-trivial solution which is regular as $r\to\infty$ and normalizable as $r\to 0$. Note that  $r\to\infty$ is the far-interior of the background which is Lifshitz, while $r\to 0$ is the boundary of the background which is AdS$_4$. From  \eqref{fPerturbed} and \eqref{gPerturbed},  the metric functions in the interior region take the form $f(r)=r^{-2z}+\cdots$ and $g(r)=\mathfrak{g}\,r^{-2}+\cdots$, where the dots represent the sub-leading terms as $r\to\infty$. Hence, the equation of motion for $\phi$ in the far-interior region reduces to  
\begin{align}\label{scalarEOMInterior}
\frac{1}{\mathfrak{g}}r^{3+z}\partial_r\left(r^{-(1+z)}\partial_r\phi\right)=L^2V'(\phi).
\end{align}
Demanding a \emph {non-trivial regular} solution in the far interior forces $\phi$ to sit at a minimum of the potential $V(\phi)$. This can easily be shown following \cite{Iqbal:2010eh}. Suppose the non-trivial regular solution in the far interior of the background behaves as $\phi=\phi_0+\delta\phi$ where $\delta\phi\sim r^\delta$ with $\delta<0$, so that $\phi\to \phi_0$ as $r\to \infty$. One then easily deduces that the left-hand side of the equation \eqref{scalarEOMInterior} vanishes, resulting in $V'(\phi_0)=0$. So, the scalar field must sit at an extremum of the potential. Now linearizing equation \eqref{scalarEOMInterior} around $\phi=\phi_0$, we find 
\begin{align}\label{LinearizedDeltaPhiEQM}
\hskip -0.07in\frac{1}{\mathfrak{g}}r^{z+3}\partial_r\left( r^{-(1+z)}\partial_r\delta\phi\right)-L^2V''(\phi_0)\delta\phi={\cal O}(\delta\phi^2).
\end{align}
The general solution of the above equation takes the form $\delta\phi(r)= a\,r^{\delta_+} +b\,r^{\delta_-}$ with 
\begin{align}
\d_\pm=\frac{1}{2}\left[(2+z)\pm\sqrt{(2+z)^2+4\mathfrak{g}L^2V''(\phi_0)}\right].
\end{align}
Because the solution $a\,r^{\delta_+}$ blows up  in the $r\to \infty$ limit, we set $a=0$.  In order for $b\,r^{\delta_-}$ to be a regular solution in the interior, one has to have $V''(\phi_0)>0$ (note that $\mathfrak{g}$ is positive). This then proves that the scalar field must sit at a minimum of $V(\phi)$, if one demands a non-trivial regular solution for $\phi(r)$ as $r\to \infty$. Thus, the regular solution takes the following form in the far-interior region of the background
\begin{align}\label{RegularPhiInterior}
\phi(r)=\phi_0+b\, r^{\delta_-}(1+\cdots), \qquad r\to \infty,
\end{align}
with 
\begin{align}\label{deltaMinus}
\d_-=\frac{1}{2}\left[(2+z)-\sqrt{(2+z)^2-8\mathfrak{g}\hskip 0.02in m^2L^2}\right].
\end{align}

Asymptotically as $r\to 0$, from \eqref{Exteriorf}, we have $f(r)=c^2r^{-2}+\cdots$ and $g(r)=r^{-2}+\cdots$. The scalar equation of motion \eqref{PhiEOM} then becomes
\begin{align}\label{PhiAsymptoticEOM}
r^4\partial_r \left(\frac{1}{r^2}\partial_r\phi\right)-\phi(\phi^2+m^2L^2)=0, \qquad r\to 0.
\end{align}
Linearizing equation \eqref{PhiAsymptoticEOM} around $\phi=0$, the general solution will be of the form
\begin{align}\label{AsymptoticPhi}
\phi= A\,r^{3-\Delta}(1+\cdots)+B\,r^\Delta (1+\cdots),
\end{align}
where $\Delta=\frac{3}{2}+\sqrt{m^2L^2+\frac{9}{4}}$. For $-\frac{9}{4}<m^2L^2<-\frac{5}{4}$, either $A$ or $B$ could be called the source and the other one the vacuum expectation value (vev) of the dual operator $\Phi$ \cite{Klebanov:1999tb}. We choose to work in the so-called  "standard quantization" where we take $A$ to be the source for $\Phi$, and $B$ as the vev of $\Phi$,  denoted by $\langle\Phi\rangle$.  (Note that for $m^2L^2\geq -\frac{5}{4}$, only the standard quantization exists.) Hence, $\Delta$ is the dimension of the neutral scalar operator $\Phi$, which is the holographic dual of the bulk neutral scalar field $\phi$.
\begin{figure}
\hskip -0.17in \includegraphics[width=70mm]{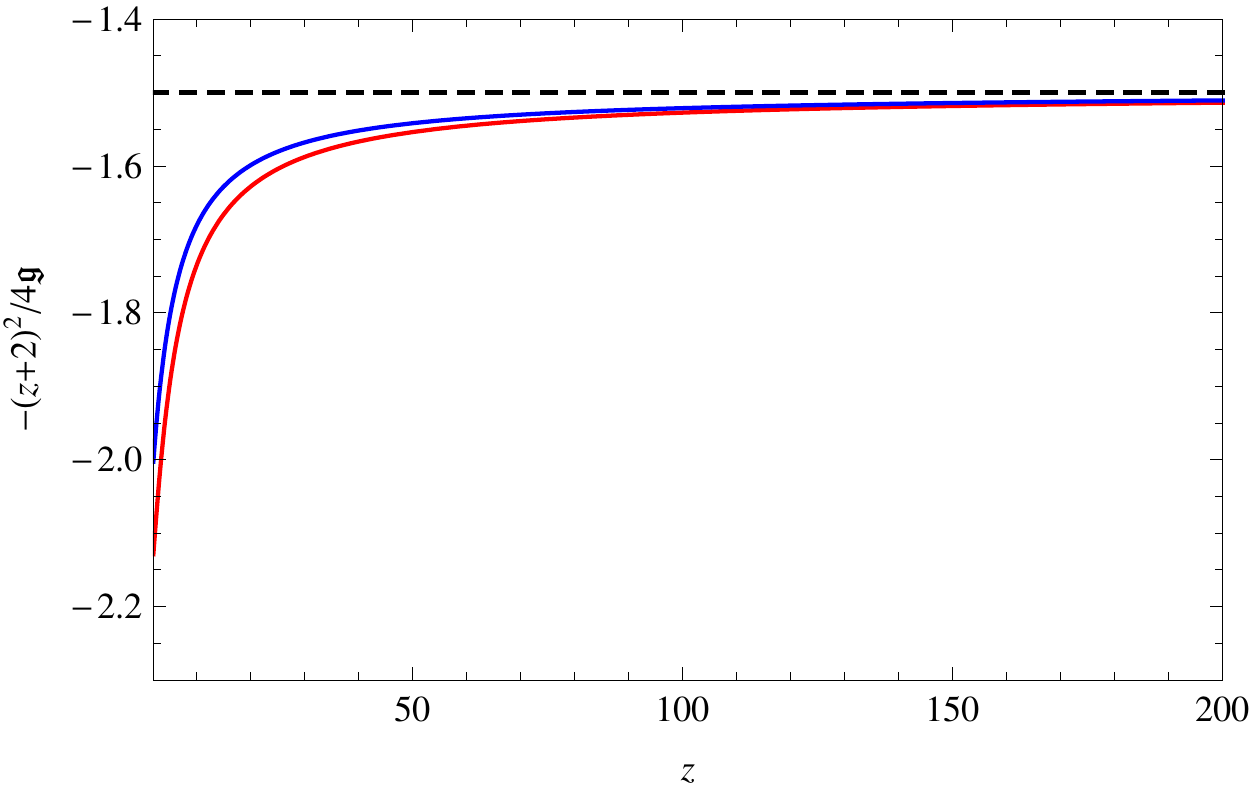}
\vskip -0.1in
\caption{\label{BFFigure} \footnotesize{Plots of $m_c^2L^2=-(z+2)^2/4\mathfrak{g}$ as function of $z$ for $m_{\rm f}=0.36$ (red) and  $m_{\rm f}=0.7$ (blue). Here $L$ is the curvature radius of the asymptotic AdS$_4$ region. As $z$ increases the curves asymptote to the dashed line, which represents the scalar BF bound in ${\rm AdS}_2$.} }
\end{figure}
Given the boundary condition  \eqref{RegularPhiInterior} for $\phi$ in the interior, for each value of $b$, the equation \eqref{PhiEOMRefined} could be numerically integrated to the asymptotic $r\to 0$ region from which both $A$ and $B$ could be determined. Repeating the process for different values of $b$, the dependence of $A$ and $B$ on $b$, and thus the dependence of $B$ on $A$, could easily be obtained. Since we are interested in the condensation of the operator $\Phi$ without turning on a source, we look for solutions where $B$ remains finite at $A=0$. 

In performing the numerics in this section, it is more convenient to define a dimensionless radial coordinate by rescaling $r \to r_s r$. As a result, we also make $b$, $A$ and $B$ dimensionless by redefining $b\to r_s^{-\delta_-}b$, $A\to r_s^{\Delta-3}A$ and $B \to r_s^{-\Delta}B$.  We also set $L=1$ in our numerics throughout this paper. Figure \ref{CondensateFigure} shows $B$ versus $A$ for the bulk neutral scalar with $m^2L^2=-2.2$ on two different electron star backgrounds, one which flows in the IR to a theory with $z=2$ and the other to a theory with $z=3$. For both of the star backgrounds, we set $m_{\rm f}=0.36$. As the plots in Figure \ref{CondensateFigure} demonstrate $B$ is non zero when $A$ is zero, indicating that the neutral operator $\Phi$ spontaneously condenses in the two boundary theories (which are the holographic duals of the two electron star backgrounds). Said another way, the aforementioned two electron star backgrounds have each formed a neutral scalar hair.

The reasoning behind the formation of a neutral scalar hair for the electron star background is similar to the RN-AdS$_4$ case discussed in \cite{Hartnoll:2008kx, Iqbal:2010eh}. Recall that the (2+1)-dimensional boundary field theory under consideration flows in the IR to a (2+1)-dimensional Lifshitz fixed-point with a finite dynamical critical exponent $z$. Let's denote the dimension of the operator $\Phi$ at the IR Lifshitz fixed-point by $\Delta_{\rm IR}$. One can then easily show that
\begin{align}\label{IRDimension}
\Delta_{\rm IR}=\frac{1}{2}\left[(2+z)+\sqrt{(2+z)^2+4\mathfrak{g}\hskip 0.02in m^2L^2}\right]. 
\end{align}
In the discussion below equation \eqref{Potential},  we have assumed that the mass square of  $\phi$ satisfies $-\frac{9}{4}<m^2L^2<0$, where the lower bound is the BF bound of the asymptotic AdS$_4$ region of the electron star background. Since for the IR Lifshitz fixed-point with $z>1$ one has  $\mathfrak{g}>1$,  there is always  a range of mass square for which the expression under the square root in \eqref{IRDimension} becomes complex, while still satisfying the BF bound of the asymptotic AdS$_4$ region. Namely, for 
\begin{align}
-\frac{9}{4}<m^2L^2<m_c^2L^2,  \,\,\,\,\,{\rm with} \,\,\,\,\,m_c^2L^2=-\frac{(z+2)^2}{4\mathfrak{g}},
\end{align}
the bulk neutral scalar field becomes tachyonic in the far-interior region  or equivalently, in the language of the the dual
field theory, the scaling dimension of $\Phi$ at the
IR Lifshitz fixed-point $\Delta_{\rm IR}$ becomes imaginary, signaling
a potential instability in the theory. Note that
$-(z+2)^2/4\mathfrak{g}$ is the scalar BF bound in the far-interior
Lifshitz geometry of the electron star background. Figure
\ref{BFFigure} shows plots of  this bound as a function of $z$ for
$m_{\rm f}=0.36$ (red) and  $m_{\rm f}=0.7$ (blue). One sees from the
plots in Figure \ref{BFFigure}  that as $z$ increases,
$m_c^2L^2$ approaches the value -3/2, which
equals the scalar BF bound in an AdS$_2$ background. (Note that $L$ is the
curvature radius of the asymptotic AdS$_4$ region.) Such behavior is
expected, and is due to the fact that  for a fixed value of $m_{\rm f}\in (0,1)$ and in the $z\to\infty$ limit, the far-interior region of the electron star background becomes ${\rm AdS}_2\times \mathbb{R}^2$  \cite{Hartnoll:2010gu}. In other words, the electron star background reverts back to the RN-AdS$_4$ solution. 

\begin{figure}
\centering
\hskip -0.05in\includegraphics[width=70mm]{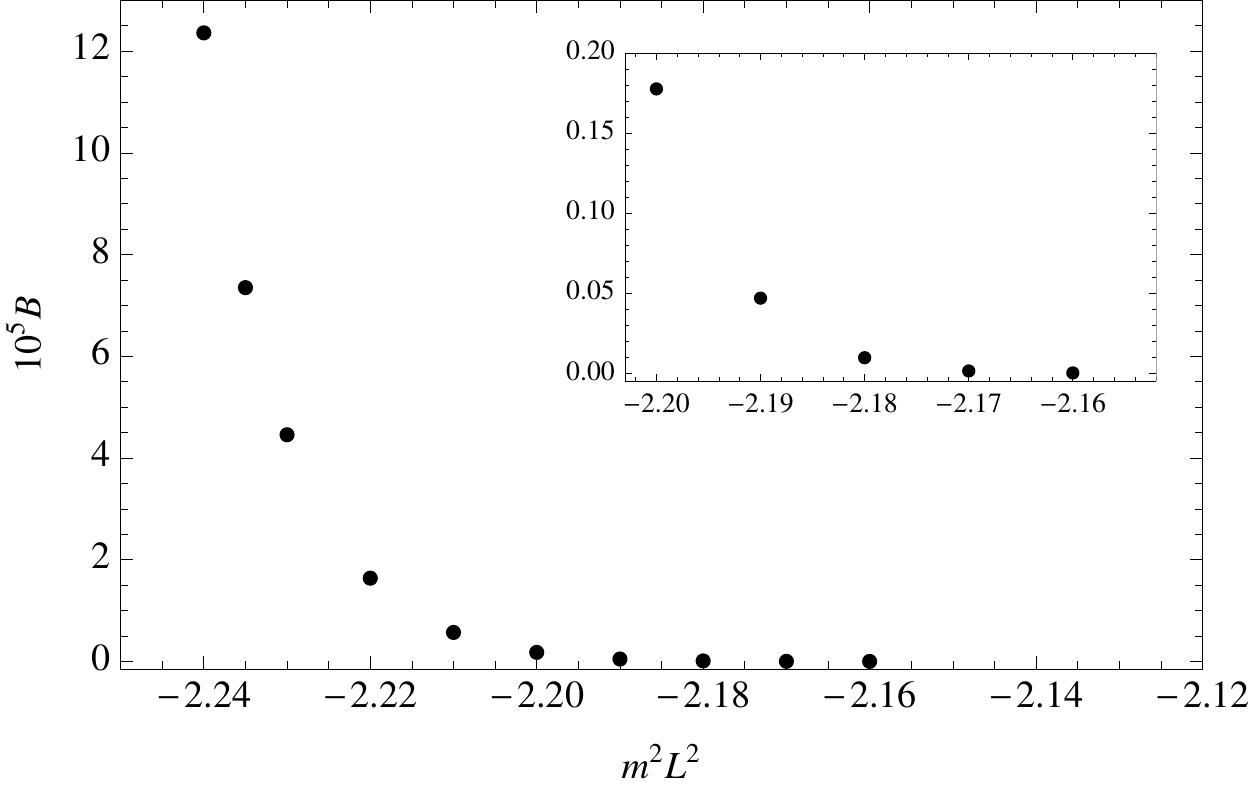}
\caption{\label{BvsMSquaredFigure} \footnotesize{$B$ versus $m^2L^2$ for the neutral scalar field $\phi$ on an electron star background with $m_{\rm f} =0.36$ and $z=2$.  
Due to numerical limitations we could not access the region of mass square very close to the BF bound, $m_c^2L^2\simeq -2.12$, of the far-interior of the star. The inset shows a close up of the behavior of $B$ for larger values of $m^2L^2$.} }
\end{figure}

In both of the plots in Figure \ref{CondensateFigure}, the mass square of the bulk scalar $m^2L^2=-2.2$  violates the BF bound of the far-interior Lifshitz region of each of the two star backgrounds while satisfying the BF bound of their asymptotic AdS$_4$
regions. Hence, we see from the plots that the neutral scalar operator
$\Phi$ spontaneously condenses in each of the two boundary
theories, and breaks the aforementioned $\mathbb{Z}_2$ symmetry. On the other hand,  if the mass square is above the BF bound in the far-interior Lifshitz region, no spontaneous condensation for $\Phi$ is
seen\footnote{Adding a double-trace deformation of the type $g\int d^3x\,\Phi^2$ (with $g<0$) to the boundary theory action, and choosing the alternative quantization where the dimension of the operator $\Phi$ satisfies $1/2<\Delta <3/2$, makes it possible for $\Phi$ to condense even if the mass square of the dual bulk neutral scalar is above the BF bound of the far-interior Lifshitz region. It would be interesting to explore, along the lines of \cite{Faulkner:2010gj},  the consequences of the aforementioned double-trace deformation for the holographic setup under consideration here. In the standard quantization, however, the mass square of the neutral scalar being below the BF bound of the far-interior Lifshitz region is the necessary and sufficient condition for the condensation of $\Phi$ in the  theory dual to the zero-temperature electron star background.} in the boundary theory, and, as a result, the $\mathbb{Z}_2$ symmetry is intact.  The plot in Figure \ref{BvsMSquaredFigure} shows the behavior of $B$ as a function of $m^2L^2$ for the bulk neutral scalar  in an electron star background with $m_{\rm f} =0.36$ and $z=2$. As the plot shows, $B$ approaches zero as $m^2L^2$ approaches $m_c^2L^2\simeq -2.12$ from below. In general, by tuning the mass square of the bulk scalar across the critical value $m_c^2L^2$, we obtain a hairy electron star background. Or, simply put in the language of the boundary theory, varying  the dimension $\Delta$ of the operator $\Phi$ across a critical dimension $\Delta_c$, the boundary theory at low energies goes through a quantum phase transition,  from an uncondensed ($\mathbb{Z}_2$-symmetric) phase, $\langle \Phi\rangle=0$, to a phase with a neutral scalar order parameter, $\langle \Phi\rangle\neq 0$, in which the $\mathbb{Z}_2$ symmetry is spontaneously broken.  Here, $\Delta_c=\frac{1}{2}\left(3+\sqrt{9-(z+2)^2/\mathfrak{g}}\right)$.

A few comments are in order. We would like to emphasize that the inequality $\mathfrak{g}>1$ plays an important role here. Indeed, should $\mathfrak{g}<1$, the BF bound in the interior Lifshitz region, for the case of $z>1$, would be less than the scalar BF bound in the asymptotic AdS$_4$ region, and, as a result, there would be no possibility for the phase transition of the kind we just described above.  

The two phases described above have the same dynamical critical
exponent $z$. This is just an artifact of the probe approximation we
have employed so far where  the backreaction of the bulk neutral scalar
field on the electron star background in neglected.  In a
subsection to follow, we will take into account such a backreaction
and show that the aforementioned zero-temperature quantum phase transition is indeed
between two phases with different $z$.  This is a particular novelty
 in certain quantum phase transitions in condensed matter
systems. Phase transitions in which the critical dynamical exponent changes across the critical point have been suggested to underlie  magnetism in the cuprates \cite{pines,sokol} and in nematic-smectic transitions \cite{fradkin}.

The existence of the neutral scalar hair is essentially goverened by
the physics in the far-interior region of the electron star
background. More concretely, in the standard quantization, it is the effective mass square of the
neutral scalar field in the Lifshitz region which determines whether
the electron star background becomes unstable to forming a neutral
scalar hair. Since the bulk scalar field is neutral, its effective mass in the
Lifshitz region is the same as the mass $m$ in the asymptotic AdS$_4$
region. So, tuning $m^2L^2$ across the BF bound in the Lifshitz
region, $m_c^2L^2=-(z+2)^2/4\mathfrak{g}$, the electron star
background undergoes a phase transition to the formation of a neutral scalar hair. Tuning the mass square of the scalar in the asymptotic region is not the only way to achieve this phase transition.  For example, a simple way for tuning  the effective mass square of the neutral scalar field in the far-interior Lifshitz region is to couple the scalar to the square of the Weyl tensor in the action \eqref{Scalar Action}, by adding to the potential $V(\phi)$ a term of, say, the following form \cite{Gubser:2005ih}
\begin{align}\label{WeylPotential}
\delta V=-\frac{1}{2} L^2l^2\phi^2 W^2,
\end{align}
where $W^2=W_{MNPQ}W^{MNPQ}$ with $W_{MNPQ}$ being the Weyl tensor. The effect of this new coupling is to make
the mass of the scalar to depend on the radial coordinate $r$. Asymptotically, $W^2\to 0$ as $r\to0$. So the mass
of the scalar in the asymptotic AdS$_4$ region is untouched by the coupling \eqref{WeylPotential}. In the far-interior region, one can easily show that $W^2\to4z^2
(-1+z)^2/(3 \mathfrak{g}^2L^4)$ as $r\to \infty$,  hence, the mass square of the scalar in the Lifshitz region is shifted by a constant
proportional to $l^2$. Thus, the effective mass square of the scalar in the Lifshitz region can be varied simply by varying $l$. The coupling $l$ could be tuned such
that the effective mass square in the Lifshitz region violates the BF bound there, in which case it is not hard to
show that the electron star background develops a neutral scalar
hair. Neither tuning $m$ nor tuning $l$ is an operation that could be
performed in a single theory. If one thinks of backgrounds such as the
electron star background, or RN-AdS$_4$, as coming from, say, M-theory, then the
couplings $m$, $l$, $\cdots$ are generally fixed. In this regard,
tuning $m$ or $l$ seems unnatural. Nevertheless, we
continue to tune $m^2L^2$ in order to be able to obtain the aforementioned
phase transition. As the transition owns its existence to what the effective
mass square of the scalar is in the far-interior region, we do expect
the qualitative features of the phase transition to stay the same
regardless of whether it has been obtained by tuning $m$, $l$, or a UV
knob (such as a magnetic field) that could be continuously changed within a single theory.  A
simple way to introduce such a knob that can be continuously varied in a single theory has been explained in \cite{Iqbal:2010eh}. 

\subsection{Free Energy of the Condensed Phase}

In this section we determine the free energy density of the boundary theory in the condensed phase $\langle\Phi\rangle\neq 0$. The free energy is given by the on-shell action, up to boundary counter terms. Assuming the bulk scalar field does not back-react on the background, the free energy of the condensed phase is just the sum of two contributions: one coming from the electron star background and the other from the neutral scalar field. 
We work in the grand canonical ensemble where the boundary theory chemical potential $\mu$ is held fixed. As shown in \cite{Hartnoll:2010gu}, the free energy density of the theory dual to the zero-temperature electron star background is given by $\Omega_1=M-\mu Q=-M/2$, with $M$ and $Q$ being the energy and charge densities, respectively.  Thus, we only need to determine the free energy density contributed by the boundary theory operator $\Phi$, denoted hereafter by $\Omega_2$.

Up to boundary counter terms, $\Omega_2$ is simply given by the action \eqref{Scalar Action} evaluated on-shell. To make the expressions less cluttered, we drop the coefficient $1/(2\kappa \lambda)$ from the action. We also set $L=1$ in this discussion. Upon integrating by parts and using the equation of motion \eqref{PhiEOM}, the action \eqref{Scalar Action} takes the form 
\begin{align}\label{NaiveOnShell}
S_{\rm o.s.}=\int d^3 x&\,\left\{\frac{1}{2}\left(\sqrt{-g}\,g^{rr}\phi\,\phi'\right)\Big|_{r=\epsilon}\right.\nonumber\\
&\left.+\frac{1}{4}\int_\epsilon^\infty dr \sqrt{-g}\,\phi^4\right\},
\end{align}
where prime denotes derivative with respect to $r$ and $\e$ is a cutoff introduced to regulate the above integral, as it is naively divergent. We will
eventually send $\epsilon$ to zero. 

If we work in an ensemble where the source is kept fixed, \ie $\delta\phi=0$, the on-shell action \eqref{NaiveOnShell} needs to be supplemented by a counter term
\begin{align}
S_{\rm c.t.}= -\frac{c}{2} (3-\Delta)\int d^3x \sqrt{-\gamma}\, \phi^2|_{r=\e},
\end{align}
to cancel the divergence. Note that  $\gamma$ is the induced metric on the boundary. Adding the above counter term to \eqref{NaiveOnShell}, and taking the $\epsilon\to 0$ limit, the renormalized on-shell action $S_{\rm ren}$ becomes
\begin{align}
S_{\rm ren}&=S_{\rm o.s.}+S_{\rm c.t.}\nonumber\\
&=V\left[\frac{c}{2}(2\Delta-3)AB+\frac{1}{4}\int^\infty_0 \hskip-0.07indr\sqrt{-g}\,\phi^4\right],
\end{align}
where
$V=\int d^3x$, and we have substituted the asymptotic expansion of the scalar field at the boundary. Since $\Omega_2=-S_{\rm ren}/V$ is the free energy density contributed by the operator $\Phi$,  the free energy density of the system $\Omega=\Omega_1 +\Omega_2$ then reads
\begin{align}
\Omega=-\frac{M}{2}-\frac{c}{2}(2\Delta-3)AB-\frac{1}{4}\int^\infty_0 \hskip-0.07in dr\sqrt{-g}\,\phi^4. 
\end{align}

If instead we work in an ensemble where the
response is fixed, \ie $\delta\phi'=0$,  then one has to add an additional counter term, analogous to the Gibbons-Hawking term \cite{Skenderis:2002wp, Herzog:2008he}, to the on-shell action in order to render the variational problem well-defined. The total counter term for this case becomes 
\begin{align}
S_{\rm c.t.}=&-\int d^3 x\,\left(\sqrt{-g}\,g^{rr}\phi\,\phi'\right)\Big|_{r=\epsilon}\nonumber\\
&+\frac{c}{2} (3-\Delta)\int d^3x \sqrt{-\gamma}\, \phi^2|_{r=\e}.  
\end{align}
Going through the same steps as before, we write the free energy density of the system as 
\begin{align}
\Omega=-\frac{M}{2}+\frac{c}{2}(2\Delta-3)AB-\frac{1}{4}\int^\infty_0 \hskip-0.07in dr\sqrt{-g}\,\phi^4. 
\end{align}

\begin{figure}
\centering
\hskip-0.1in\includegraphics[width=70mm]{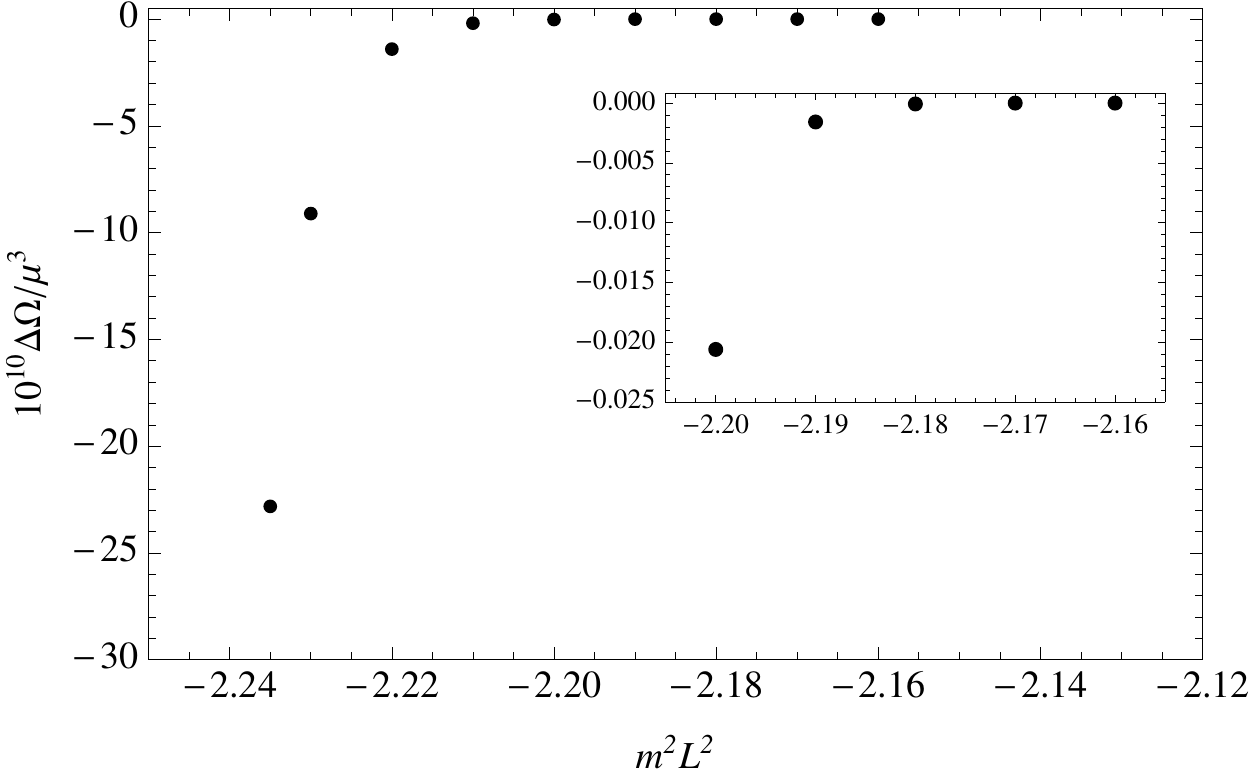}
\vskip -0.1in
\caption{\label{OmegavsM} \footnotesize{Plot of $\Delta\Omega/\mu^3$ versus $m^2L^2$.  For the background, we set $z=2$ and $m_{\rm f}=0.36$. Numerical difficulties prevented us from obtaining more data points for the values of $m^2L^2$ very close to the critical value $m_c^2L^2\simeq -2.12$.}} 
\end{figure}

The condensed phase, $\langle\Phi\rangle\neq0$,  corresponds, in the standard quantization, to $A=0$ and $B\neq0$ while the uncondensed phase is given by $A=B=0$. Indeed, in the latter phase,  there is only the trivial solution $\phi(r)=0$  in the bulk. The change in free energy is then given by 
\begin{align}
\Omega_{\rm condensed}-\Omega_{\rm uncondensed} = -\frac{1}{4}\int^\infty_0 \hskip-0.07in dr\sqrt{-g}\,\phi^4.
\end{align}
Therefore, we see that when the operator $\Phi$ condenses, the free energy of the system is actually lower than that of the uncondensed phase. Figure \ref{OmegavsM} shows a plot of the change in free energy density, $\Delta\Omega=\Omega_{\rm condensed}-\Omega_{\rm uncondensed}$, versus $m^2L^2$. Clearly, the condensed phase has a lower free energy than the uncondensed phase, the difference approaches zero when the critical point is approached from below.

\subsection{Zero-Temerature Phase Transition}

In this subsection we analyze the nature of this zero-temperature phase transition and argue that it has a BKT character, meaning that close to the critical point in the condensed phase $\langle \Phi\rangle \sim \mu^\Delta \exp(-{c}_1/\sqrt{\nu_c-\nu})$, where $\nu$ is a parameter that can be tuned across the critical point and $c_1>0$.  

Consider the boundary theory dual to the AdS$_{d+1}$ background, with a UV scale.  For this theory, the authors of \cite{Kaplan:2009kr} have argued on general grounds  that if the mass square of a bulk scalar is slightly below the BF bound, an IR scale $\Lambda_{\rm IR}$ will be generated, which is related to the UV scale $\Lambda_{\rm UV}$  via $\Lambda_{\rm IR}=\Lambda_{\rm UV} \,\exp(-c_2/\sqrt{\nu_c-\nu})$ with $c_2 >0$. Concrete examples of this phenomenon were constructed in \cite{Jensen:2010ga, Iqbal:2010eh} using a background with an AdS$_2$ near horizon geometry. Also, it was argued in \cite{Jensen:2010ga} that the same phenomenon would occur in a boundary theory dual to a Lifshitz-like background \cite{Jensen:2010ga}. As shown below,  the discussion in \cite{Kaplan:2009kr, Iqbal:2010eh} for a BKT-type phase transition can handily be generalized to our case as well. 

Consider the equation of motion for $\phi$ given by equation \eqref{scalarEOMInterior}.  Linearizing  around $\phi=0$ in the far-interior Lifshitz region of the electron star background, this equation can be put in the form 
\begin{align}
\chi''(r)
=\frac{\mathfrak{g}}{r^2}\left[-r^{2z}\omega^2+\left(m^2-m_c^2\right)L^2-\frac{1}{4\mathfrak{g}}\right]\chi(r),
\end{align}
where we assumed a time-dependence of $e^{-i\omega t}$ for $\phi$, and defined $\phi(r)=r^{(z+1)/2}\chi(r)$. Note that in the above equation $m_c^2L^2=-(2+z)^2/4\mathfrak{g}$ is the BF bound of the far-interior
Lifshitz region. For the mass square below the BF bound, there will
exist a negative-energy bound state if $0<r<\infty$. On the other
hand, if the geometry is cut by some suitably chosen UV and IR
``walls",  denoted by $r_{\rm UV}$ and $r_{\rm IR}$  such that $r_{\rm
  IR}\geq r\geq r_{\rm UV}$, the existence of negative-energy bound
states could be prevented. This is actually the case in our setup. The
far-interior Lifshitz region does not extend all the way to the
boundary, and at some radius matches to the exterior geometry.  Also,
the IR wall is provided by the condensate, whose existence is
essentially due to the stabilizing effect of the quartic term in 
the scalar potential \eqref{Potential}. 

Since we are interested in the onset of an instability we look at the $\omega=0$ solutions. As argued in \cite{Kaplan:2009kr, Iqbal:2010eh} the exact form of the boundary conditions for $\phi$ at the walls are not important and could be taken to be Dirichlet. So, we choose $\phi(r_{\rm UV})=\phi(r_{\rm IR})=0$. The zero-energy solution satisfying the Dirichlet boundary condition at $r_{\rm UV}$ is found to be \cite{Kaplan:2009kr, Iqbal:2010eh}
\begin{align}
\phi(r)=r^{(z+2)/2}\sin\left[\sqrt{\mathfrak{g}\left(m_c^2L^2-m^2L^2\right)}\log\frac{r}{r_{\rm UV}}\right].
\end{align}
To satisfy the boundary condition at $r_{\rm IR}$, one should then have  
\begin{align}\label{rIR}
r_{\rm IR}=r_{\rm UV} \exp\left[\frac{\pi}{\sqrt{\mathfrak{g}\left(m_c^2L^2-m^2L^2\right)}}\right].
\end{align}
Note that $r_{\rm IR}/r_{\rm UV}= \Lambda_{\rm UV}/\Lambda_{\rm IR}$. For $r_{\rm IR}$ given by the above formula, the ground state has zero energy. So, there will be no negative-energy bound states. On the other hand, for larger $r_{\rm IR}$ there will be a negative-energy ground state, hence instability.  For the mass square of the dual bulk scalar slightly less than the BF bound, the IR Lifshitz fixed-point is still scale invariant over a large energy scale. For energies below $\Lambda_{\rm IR}$, the operator $\Phi$ wants to condense and end the instability.  For the mass square slightly less than the BF bound, the IR (mass) dimension of $\Phi$ according to \eqref{IRDimension} is almost $(z+2)/2$. So, following the argument given in \cite{Iqbal:2010eh}, and given that $\Lambda_{\rm UV}\sim \mu$, one then concludes 
\begin{align}\label{BKTCondensate}
\langle\Phi\rangle\sim \mu^\Delta \exp\left[-\frac{(z+2)\pi}{2\sqrt{\mathfrak{g}\left(m_c^2L^2-m^2L^2\right)}}\right].
\end{align}  
Note that there are also an infinite number of states, characteristic
of Efimov states \cite{Efimov, Jensen:2010ga,Iqbal:2010eh}, with
$\langle\Phi\rangle_n \sim\mu^\Delta\exp\left[-(1+z/2)n\pi/\sqrt{\mathfrak{g}\left(m_c^2L^2-m^2L^2\right)}\right]$,
where $n=2,3,4,\cdots$. The ground state, however, is given by
\eqref{BKTCondensate}. The scaling of the condensate as in
\eqref{BKTCondensate} signifies that the underlying transition
is of the BKT-type. 

\subsection{Backreaction}\label{Backreaction}

In this section, we consider the backreaction of the neutral scalar field on the electron star background, and show that if the mass square of the scalar is below the BF bound of the far-interior region, the dual operator in the boundary theory condenses. Moreover, in the condensed phase the value of the dynamical critical exponent $z$ is different from that of the uncondensed phase. 

To consider the effect of backreaction of the scalar field  on
the background, we need to modify the expression for  $T_{MN}$ in \eqref{Stress} by including the energy-momentum tensor of the scalar. Note that since the scalar field is neutral, the current $J_N$ is unchanged. The ansatz that we take for the metric and the gauge field is the same as in \eqref{BackgroundAnsatz} and \eqref{GaugeAnsatz}, respectively. For the scalar, as in the previous sections, we take the ansatz $\phi=\phi(r)$. Substituting the ansatz into the Einstein-Maxwell-fluid-scalar equations of motion, one obtains
\begin{widetext}
\begin{align}
0&=\frac{1}{r}\left(\frac{f'(r)}{f(r)}+\frac{g'(r)}{g(r)}+\frac{4}{r}\right) +\left[p(r)+\rho(r)\right] g(r)+\frac{1}{2\lambda}\phi'(r)^2,\label{BackEq1}\\  
0&=p'(r)+\frac{1}{2}\frac{f'(r)}{f(r)}\left[p(r)+\rho(r)\right]-\frac{h'(r)}{\sqrt{f(r)}}\,\sigma(r),\label{BackEq2}\\ 
0&=\frac{f'(r)}{rf(r)}-\frac{h'(r)^2}{2f(r)}+\left[3+p(r)\right]g(r)-\frac{1}{r^2}+\frac{1}{4\lambda}\phi'(r)^2-\frac{g(r)}{2\lambda}L^2V(\phi),\label{BackEq3}\\ 
0&= h''\hskip-0.02in(r)\hskip-0.02in+\hskip-0.02in\frac{r}{2}\left[p(r)+\rho(r)\right]g(r)h'(r)\hskip-0.02in-\hskip-0.03in g(r)\sqrt{f(r)} \sigma(r)+\frac{1}{4\lambda}rh'(r)\phi'(r)^2,\label{BackEq4}\\ 
0&=\phi''(r)+\frac{1}{2}\left(\frac{f'(r)}{f(r)}-\frac{g'(r)}{g(r)}-\frac{4}{r}\right)\phi'(r)-g(r)L^2V'(\phi).\label{BackEq5}
\end{align}
\end{widetext}

As before, to solve the above equations, we need to specify the
equation of state for the fermion fluid. Compared to the case without the scalar backreaction, the equation \eqref{BackEq2} is unchanged . This is due to the fact that in our model, the neutral scalar does not couple
directly to the fermion fluid. Thus, the fermion density of states is given
by \eqref{DensityOfStates}. Also, similar to  the case without the
scalar backreaction, we adopt a locally flat approximation where the
fermion dynamics is determined by the local chemical potential, whose
form is given in \eqref{LocalChemicalPotential}. As a result, the
formulae for $\sigma$, $\rho$, and $p$ are the same as before,  with the understanding that the metric and the gauge field functions in these expressions are now the backreacted ones. We do not write the formulae for $\sigma$, $\rho$, and $p$ here as they can be found in \cite{Hartnoll:2010gu}. It can easily be verified that the expressions for $\sigma$, $\rho$, $p$ satisfy the equation \eqref{BackEq2}.  Thus, in order to determine the backreacted background, one has to solve the remaining equations, namely  the equations   \eqref{BackEq1}, \eqref{BackEq3}, \eqref{BackEq4} and \eqref{BackEq5}.

In the far-interior region $r\to\infty$, we anticipate a Lifshitz-type geometry.  This is due to the fact that the local charge density screens the electric field making it massive, and massive vector fields usually give rise to Lifshitz geometries \cite{Kachru:2008yh}.  Indeed, in the region $r\to \infty$, the following  
\begin{align}
\frac{ds^2}{L^2}&=-\frac{dt^2}{r^{2z}}+\frac{1}{r^2}\left(dx^2+dy^2\right)+\frac{\mathfrak{g}}{r^2}dr^2,\nonumber \\ \label{IRBackReactedSol}
A&=\frac{eL}{\kappa} \frac{\mathfrak{h}}{r^z}dt, \qquad \qquad\phi=\phi_0,
\end{align}
is an exact solution to the equations of motion, with $\phi_0=0, \pm
\sqrt{-m^2L^2}$.  The expressions for $\mathfrak{g}$ and
$\mathfrak{h}$ are identical to the ones given in \eqref{GandH} except
that, with the scalar backreaction included, and when $\phi_0=\pm
\sqrt{-m^2L^2}$, the dynamical exponent $z$ acquires contributions from two more parameters $mL$ and $\lambda$, namely, $z=z(m_{\rm f}, \beta, mL, \lambda)$. Note that for $\phi_0=0$, $z$ does not depend on either $mL$ or $\lambda$. For example, without the scalar backreaction, $z=2$ for $m_{\rm f}=0.36$ and $\beta=19.95$.  Including the scalar backreaction, $z$ is modified to approximately 2.025  for $m^2L^2=-2.2$, $\lambda=10$ and the same values of $m_{\rm f}$ and $\beta$. 

Since the far-interior solution is of the Lifshitz type, following
essentially the same argument as before, one can analyze equation
\eqref{BackEq5}  to show that demanding a non-trivial regular solution
for $\phi(r\to\infty)$ requires the scalar to sit at a minimum of
$V(\phi)$. This is exactly the case when $z$ acquires contributions from $mL$ and $\lambda$.  The form of this non-trivial regular solution in the far-interior region is simply given by \eqref{RegularPhiInterior}, with $z$ and $\mathfrak{g}$ given by their new backreacted values.  

In order to  flow up to the asymptotic region, we perturb away from the far-interior Lifshitz solution \eqref{IRBackReactedSol} by taking the ansatz 
\begin{align}
f(r)&=\frac{1}{r^{2z}}\left(1+\mathfrak{f}_1 r^\alpha+\cdots\right),\\
g(r)&=\frac{\mathfrak{g}}{r^{2}}\left(1+\mathfrak{g}_1 r^\alpha+\cdots\right),\\
h(r)&=\frac{\mathfrak{h}}{r^{z}}\left(1+\mathfrak{h}_1 r^\alpha+\cdots\right) ,\\
\phi(r)&=\phi_0+b\, r^{\delta}+\cdots.\label{BackPhiInterior}
\end{align}
We demand the perturbation to grow as $r\to 0$ and not blow up as $r\to\infty$. We also discard the perturbation which gives rise to turning on a finite temperature. Substituting the above ansatz back into the equations of motion, and taking into account the above consideration for the type of perturbation, we then find that the expressions for $\mathfrak{f}_1$, $\mathfrak{g}_1$, $\mathfrak{h}_1$ and $\alpha$ take the same form as in the case with no scalar backreaction, although their numerical values are changed due to the change in $z$. Similarly, the expression for $\delta$ is the same as $\delta_-$ in \eqref{deltaMinus}. Thus, similar to the case without backreaction, the scalar field sits at a minimum of $V(\phi)$, which results in $z$ acquiring a contribution from $mL$ and $\lambda$.  The perturbation described above can then be thought of as an irrelevant deformation of the (2+1)-dimensional IR Lifshitz fixed-point, which is assumed to be the holographic dual of the bulk solution \eqref{IRBackReactedSol} with $\phi_0$ siting at the bottom of the potential.

Note that the backreacted geometry is asymptotically  AdS$_4$, and
the leading-order expansion of $\phi$ (expanded around $\phi=0$) in
that region is the same as in \eqref{AsymptoticPhi}. Thus, if the
non-trivial regular solution for $\phi$ in the far-interior region
could be matched to a normalizable solution in the asymptotic region,
then the boundary theory dual to this background would be in a
condensed phase where the dynamical exponent $z$ is different from the value when the boundary theory is in an uncondensed phase. So, one can by tuning a knob, which, for simplicity, could be taken to be the conformal dimension of the operator $\Phi$, drive the boundary theory from an uncondensed phase,  $\langle\Phi\rangle=0$, with $z=z_1$ to a condensed phase, $\langle\Phi\rangle\neq0$, where $z=z_2\neq z_1$. 

\begin{figure}
\centering
 \hskip-0.1in\includegraphics[width=70mm]{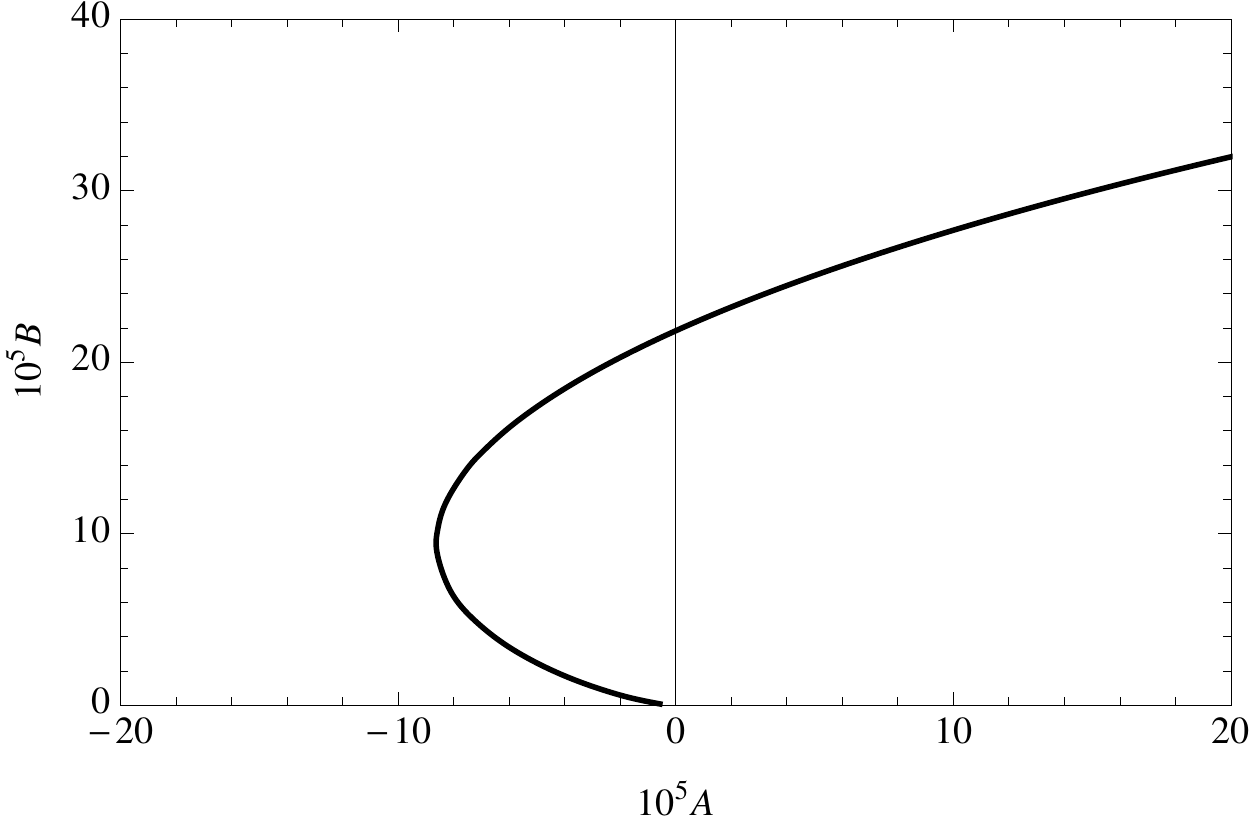}
\vskip -0.1in
\caption{\label{BvsABackReacted} \footnotesize{Plot of $B$ versus $A$ for  $\lambda=1000$. Here $z=4$, $m_{\rm{f}}=0.36$ and $m^2L^2=-2.2$. The BF bound of the far-interior region is $m_c^2L^2\simeq-1.952$.}} 
\end{figure}

In order to find a normalizable solution for $\phi$ in the asymptotic region, we scan different values of $b$ (or, more precisely, the scale-invariant combination $b/\mathfrak{f}_1^{\delta/\alpha}$) to determine whether there is any solution for which $A=0$ while $B\neq0$. Figure \ref{BvsABackReacted} shows a  plot of $B$ against $A$ for the case of $z=4$ and $\lambda=1000$. As the plot shows,  the qualitative behavior of $B$ against $A$ is similar to the case when the scalar backreaction is ignored. The plots show that the boundary theory dual to the  backreacted background is in a phase where $\langle\Phi\rangle\neq0$.

\section{Neutral Scalar Condensation at Finite Temperature}\label{SectionFour}

In this section we extend our studies of the condensation of the operator $\Phi$ to finite temperature. The boundary field
theory in this case is dual to the finite-temperature electron star
background \cite{Puletti:2010de, Hartnoll:2010ik}. As we alluded to
earlier, a finite-temperature electron star background exists only for
temperatures less than a critical temperature $T_*$. For $T>T_*$, the background is a non-extremal RN-AdS$_4$. The
condensation of a neutral scalar operator for the theory dual to a
non-extremal RN-AdS$_4$ background was already addressed in \cite{Iqbal:2010eh}.  Hence, the temperatures we consider in this section are always in the range $T<T_*$ for which the electron-star background is the preferred solution.

The operator $\Phi$ is dual to the neutral scalar field $\phi$ whose bulk action is given in \eqref{Scalar Action}. The assumption regarding the mass $m$ of the scalar field is as before. Also, we first consider the case where the scalar backreaction is ignored, namely, for large $\lambda$. The equation of motion for the scalar field is given in \eqref{PhiEOMRefined}. 
We are after a non-trivial solution which is regular as $r\to r_0$ and normalizable as $r\to 0$. Such a solution in the bulk characterizes a phase of the boundary theory where the dual operator $\Phi$ condenses. At finite temperature, demanding a non-trivial regular solution for the scalar $\phi$ close to the horizon does not require the scalar to be sitting at the bottom of the potential. This can easily be verified by expanding equation \eqref{PhiEOMRefined} close to the horizon. Note that close to the horizon, in the so-called inner region of the finite-temperature electron star background, the metric functions are given by \eqref{Inner Region}. Indeed, once expanded close to the horizon, equation \eqref{PhiEOMRefined} admits a regular solution of the form 
\begin{align}\nonumber
\phi(r)&=\phi_0+\phi_1(r_0-r)+\phi_2(r-r_0)^2+\cdots,
\end{align}
where $\phi_1$, $\phi_2$ and all other coefficients are determined in terms  of $\phi_0$. 

\begin{figure*}
\centering
 \includegraphics[width=65mm]{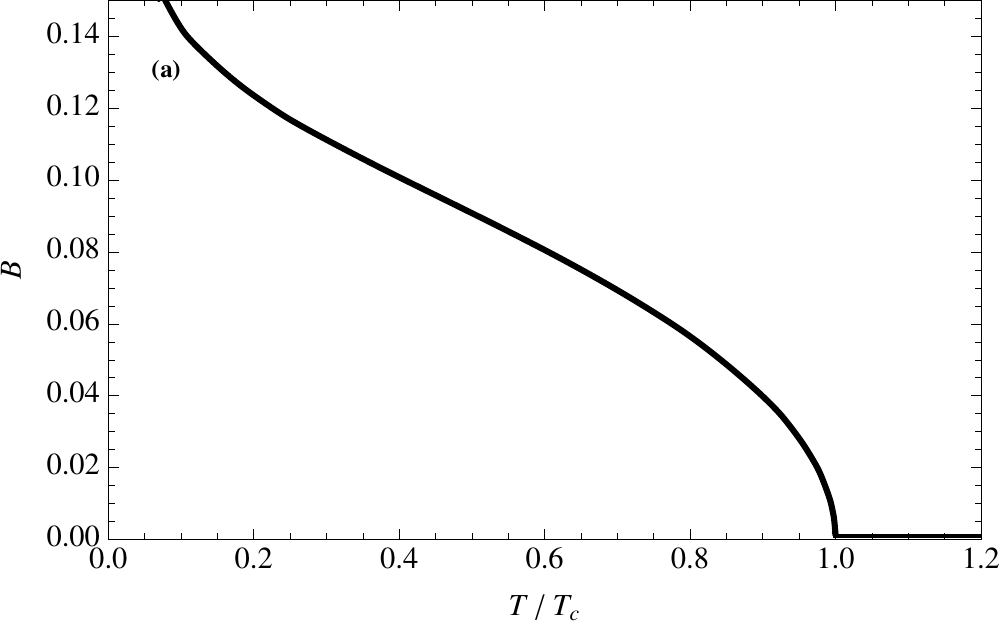} \qquad\qquad
\includegraphics[width=65mm]{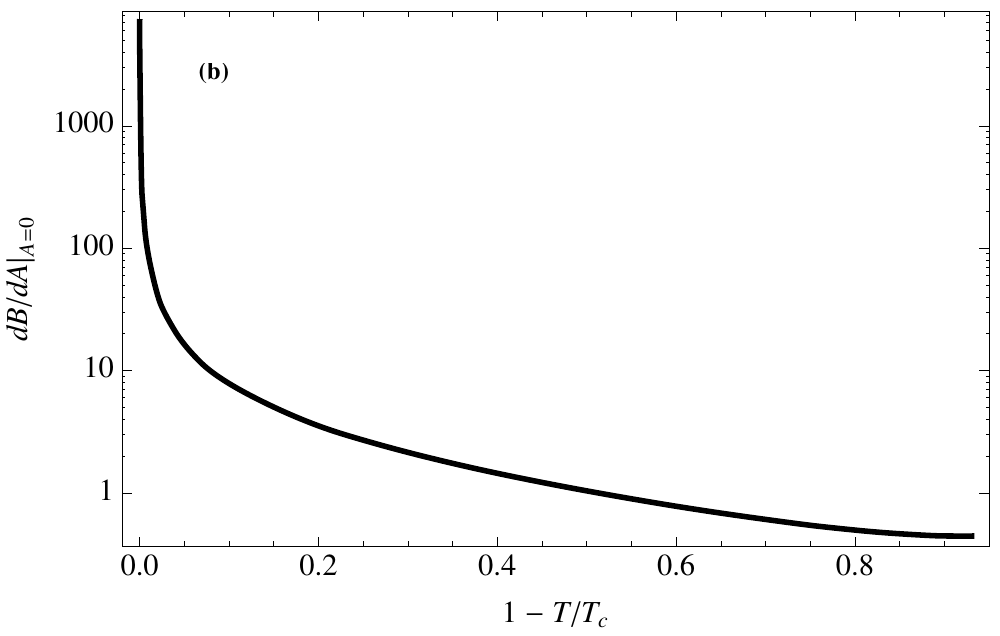}\\\vskip 0.15in
\includegraphics[width=65mm]{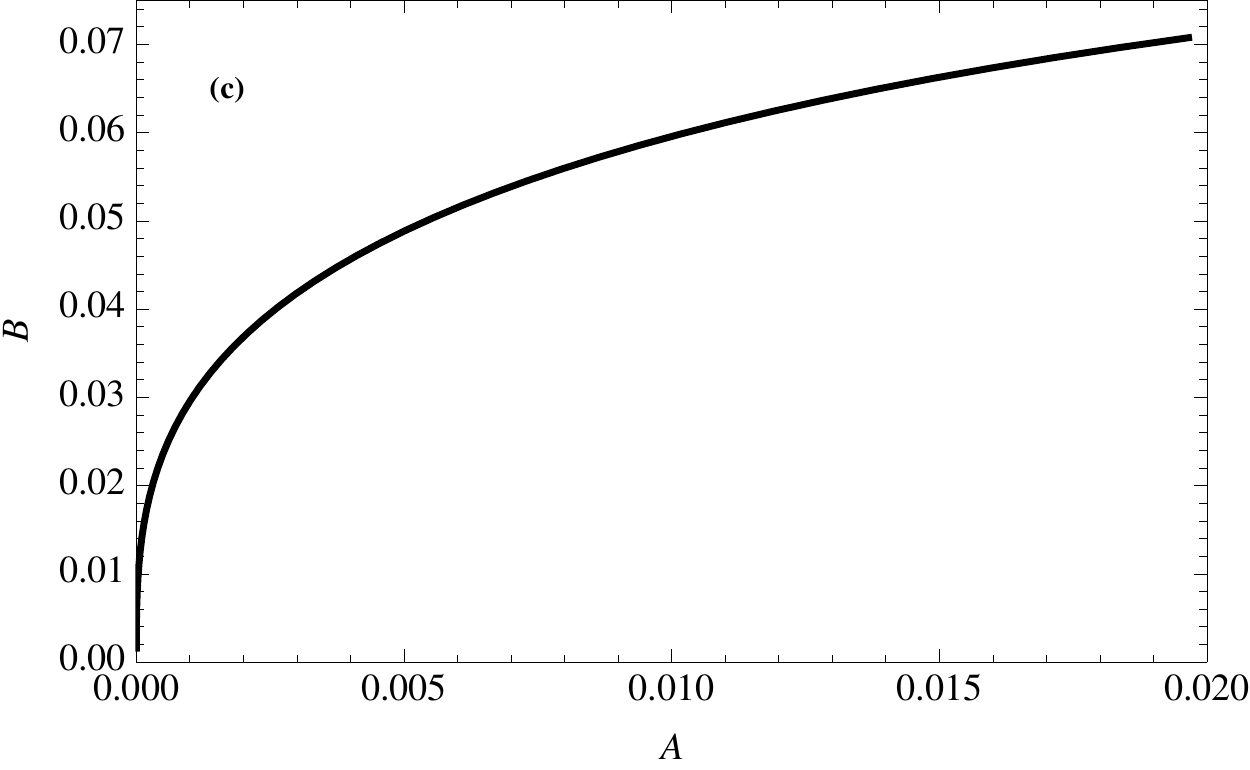} \qquad\qquad
\includegraphics[width=65mm]{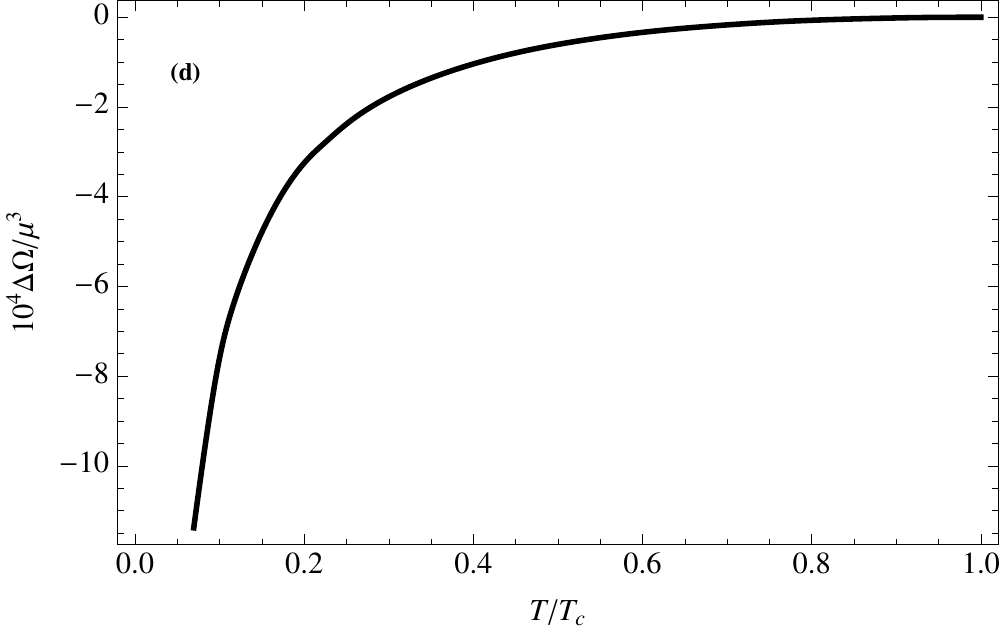}
\vskip -0.1in
\caption{\label{CriExp} \footnotesize{Plots of (a) $B$ versus $T/T_c$, (b) $dB/dA$ at $A=0$ as function of $1-T/T_c$  (c) $B$ versus $A$ at $T=T_c$ and (d) $\Delta\Omega$ versus $T/T_c$.  We set $m^2L^2=-2.2$ and for the background,  $m=0.7$ and $\b=10$. From these plots, the critical exponents $\b_c$, $\gamma_c$, $\d_c$, and $\nu_c$ are found to be $\b_c\simeq 0.506$, $\gamma_c\simeq 1.015$, $\d_c\simeq 3.07$ and $\nu_c\simeq 2.061$ which indicate a second-order phase transition with mean field exponents.} }
\end{figure*}

The asymptotic behavior of $\phi$
(expanded around $\phi=0$) will be given by the same expression as in
\eqref{AsymptoticPhi}, as both the zero- and finite-temperature electron star backgrounds
are asymptotically AdS$_4$. As in the zero-temperature case in the previous
section, we choose to work in the standard quantization where $A$ is
the source and $B$ is the vev of the dual operator
$\Phi$. Thus, by varying $\phi_0$,  one can search for a normalizable
solution asymptotically with $A=0$ and  $B\neq0$.  For numerical
computations in this section, we define a dimensionless radial
coordinate by rescaling $r \to r_0 r$, and also work with
dimensionless coefficients $A$ and $B$ defined through $A\to
r_0^{\Delta-3}A$ and $B \to r_0^{-\Delta}B$. 
\begin{figure}
\centering
\hskip-0.15in \includegraphics[width=73mm]{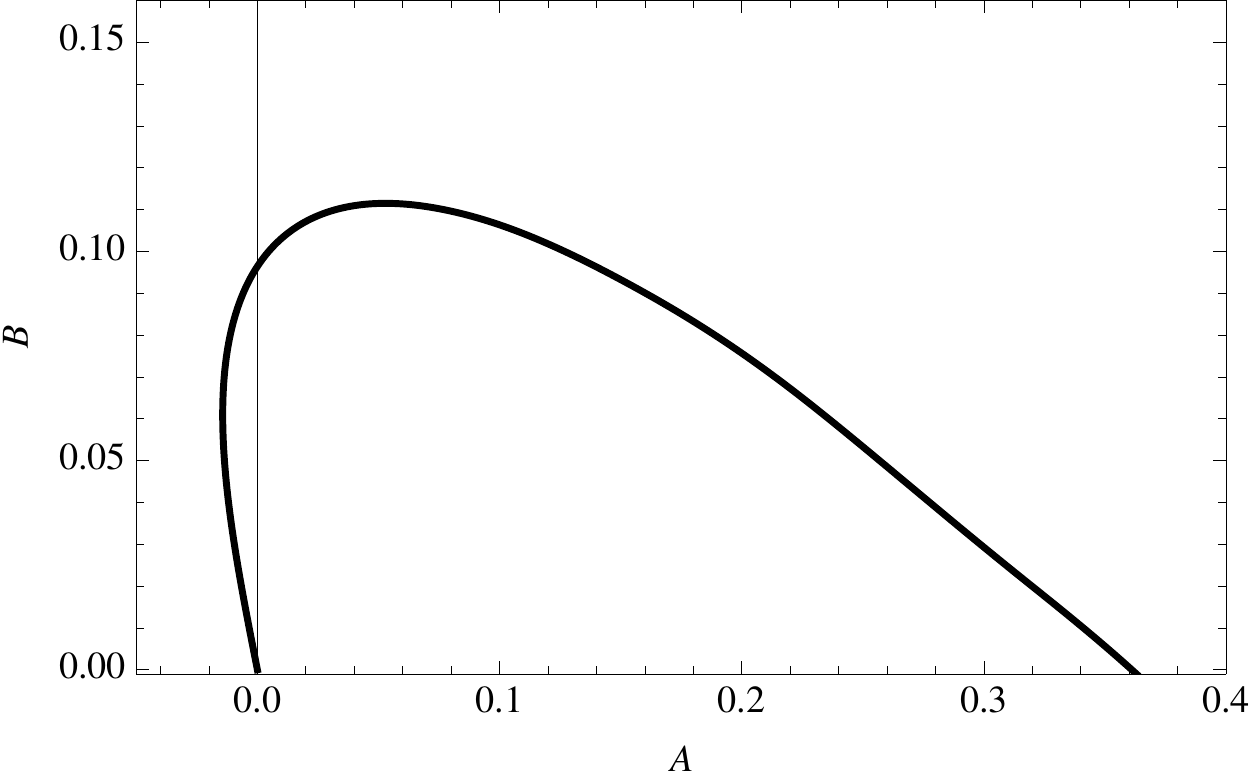}
\vskip -0.1in
\caption{\label{BvsABackFiniteT} \footnotesize{Plot of $B$ versus $A$ for $m^2L^2=-2.2$ and $T/T_c\simeq0.443$. For the background, we set $m_{\rm f}=0.7$ and $\beta=10$.}} 
\end{figure}
\begin{figure}
\centering
\hskip-0.23in\includegraphics[width=74mm]{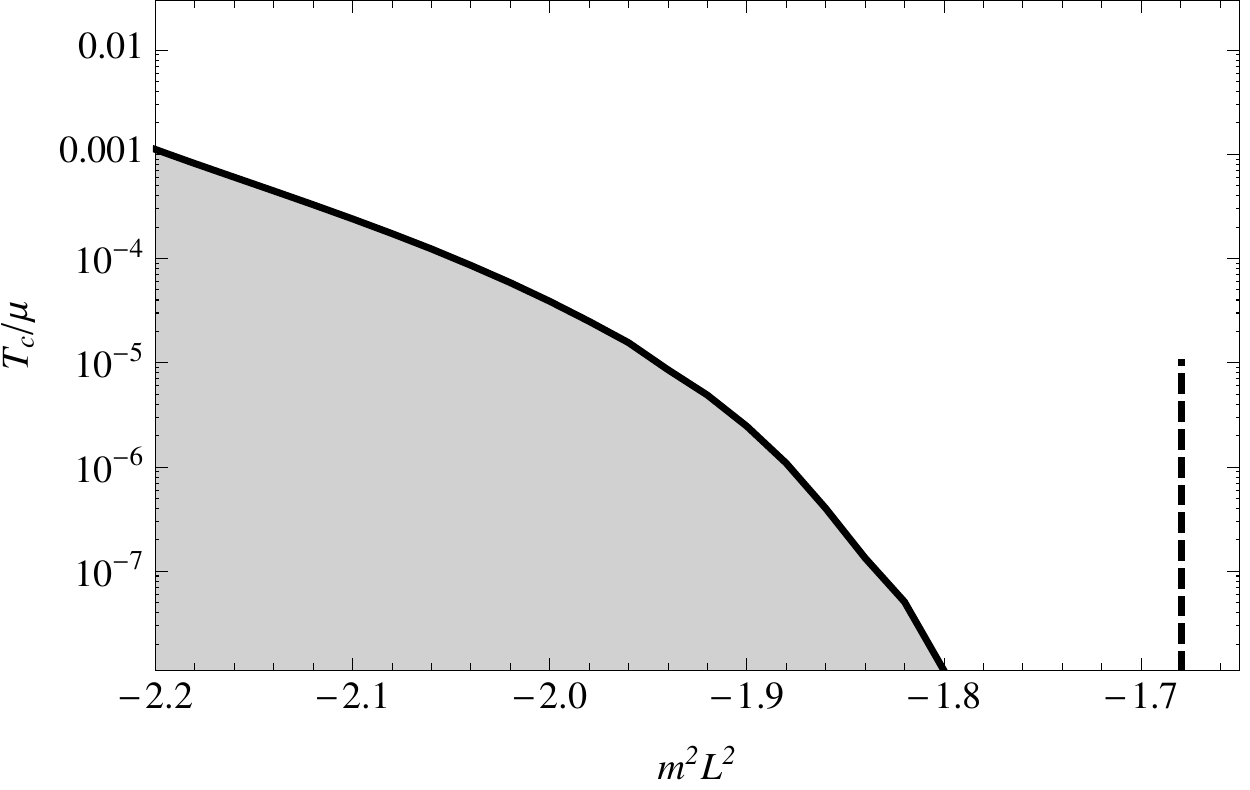}
\vskip -0.1in
\caption{\label{TcvsMSquared} \footnotesize{$T_c$ versus $m^2L^2$.  For the background, we set $m_{\rm f}=0.7$ and $\beta=10$. The shaded region represent the condensed phase, and the dashed line marks the BF bound of the far-interior region as $T\to0$.}} 
\end{figure}
Our numerics show that the operator $\Phi$ condenses below some critical temperature $T_c$ as long as the mass squared of the scalar field satisfies $-9/4<m^2L^2\leq m_c^2L^2$. Note that the lower limit is there to ensure stability in the asymptotic AdS$_4$ region. Figure
\ref{BvsABackFiniteT} shows a plot of $B$ versus $A$ for $m^2L^2=-2.2$
on an electron star background with $m_{\rm
  f}=0.7$ and $\beta =10$. For the plot,  $T/T_c\simeq 0.443$. Notice that as $\phi_0\to
-\phi_0$, $A\to -A$ and $B\to -B$. So, without loss of generality, we
have just plotted half the curve of $B$ versus $A$ in Figure
\ref{BvsABackFiniteT}, the half for which  $\phi_0$ is
non-negative. Our numerics also show that $T_c \to 0$ as $m^2L^2$
approaches the critical value from below, as shown in Figure
\ref{TcvsMSquared}, although, due to numerical difficulties, we could
not probe larger values of $m^2L^2$, namely the ones very close to
$m_c^2L^2$.  Moreover, as shown in the plots of Figure \ref{CriExp},
we find that close to $T_c$,  $B\sim(1-T/T_c)^{\b_c}$ with $\b_c\simeq
0.506$, $dB/dA|_{A=0}\sim(1-T/T_c)^{-\gamma_c}$ with $\gamma_c=1.015$,
and at $T=T_c$, $B\sim A^{1/\d_c}$ with $\d_c\simeq3.07$. Also,
Figure \ref{CriExp}(d) shows a plot of the change in the free energy density $\Delta\Omega=\Omega_{\rm condensed}-\Omega_{\rm uncondensed}$
versus the temperature. It shows that whenever there is a condensed
phase, its free energy is always lowered when compared to the
uncondensed phase. Hence, the condensed phase is energetically favored
over the uncondensed phase. As  temperature approaches $T_c$ from below, the difference in free energy approaches zero as  $\Delta\Omega\sim (1-T/T_c)^{\nu_c}$, with the numerically obtained value of $\nu_c\simeq 2.061$. 

The numerically obtained
critical exponents $\beta_c$, $\gamma_c$, $\delta_c$ and $\nu_c$ are indicative of a second-order phase transition with mean field exponents. The finite-temperature phase transition obtained here is indeed similar to the result found in \cite{Iqbal:2010eh} for the scalar condensation in the boundary theory dual to the non-extremal RN-AdS$_4$ background. This is perhaps not surprising as the nature of this phase transition at  finite temperature could be traced back to the analyticity of $A$ and $B$ as a function of $\phi_0$ close to the horizon \cite{Iqbal:2010eh}. 

\subsection{Backreaction at Finite Temperature}

We now take into account  the backreaction of the neutral scalar field
on the finite-temperature electron star background. Again,  since the
neutral scalar does not couple directly to the fermion, the fermion
density of states is given by \eqref{DensityOfStates}. This assumes
that the effect of the temperature on the fermionic density of states is negligible. We also assume that  the fermion dynamics are determined by the local chemical potential, with a form given in \eqref{LocalChemicalPotential}, with the same formulae for $\sigma$, $\rho$, $p$ given as in the case without scalar backreaction. Similar to the case without scalar backreaction, the finite-temperature background is divided into three regions. The radii  $r_1$ and $r_2$ which separate the three different regions  are solutions to the equation  $\mu_{\rm loc}(r)=m_{\rm f}$. Our discussion of the backreacted background in the following will be brief as its characteristics are very much similar to the case without the backreaction of the neutral scalar.  In the inner region, $r_0\geq r\geq r_2$, the solution is obtained by solving the equations \eqref{BackEq1}, \eqref{BackEq3}, \eqref{BackEq4} and \eqref{BackEq5} with $p=\rho=\sigma=0$, namely 
\begin{align}
0&=\frac{1}{r}\left(\frac{f'(r)}{f(r)}+\frac{g'(r)}{g(r)}+\frac{4}{r}\right)+\frac{1}{2\lambda}\phi'(r)^2,\label{EintBackReacted2}\\
0&=\frac{f'(r)}{rf(r)}-\frac{h'(r)^2}{2f(r)}+\left[3-\frac{1}{2\lambda}L^2V(\phi)\right]g(r)\nonumber\\
&-\frac{1}{r^2}+\frac{1}{4\lambda}\phi'(r)^2,\\
0&=h''(r)+\frac{1}{4\lambda}rh'(r)\phi'(r)^2,\\
0&=\phi''(r)+\frac{1}{2}\left(\frac{f'(r)}{f(r)}-\frac{g'(r)}{g(r)}-\frac{4}{r}\right)\phi'(r)\nonumber\\
&-g(r)L^2V'(\phi).\label{PhiEOSBackReacted2}
\end{align}
Our objective in this subsection is to study the condensation of the operator $\Phi$ in the boundary theory dual to the backreacted finite-temperature electron-star background.  This means that we are after a solution for $\phi$ which is regular at the horizon (as well as normalizable in the asymptotic region). So, we start by taking the following ansatz for the expansion of the  solution near the horizon 
\begin{align}
g(r)&=\frac{g_0}{r-r_0}+g_1+\cdots,\\
f(r)&=f_0(r-r_0)+f_1(r-r_0)^2+\cdots,\\
h(r)&=h_0(r-r_0)+h_1(r-r_0)^2+\cdots,\\
\phi(r)&=\phi_0+\phi_1(r-r_0)+\phi_2(r-r_0)^2+\cdots.
\end{align}
It is more convenient to rescale the radial coordinate such that the horizon is at  $r_0=1$. Substituting the ansatz back into the equations \eqref{EintBackReacted2}-\eqref{PhiEOSBackReacted2}, one finds that all of the coefficients  in the above expansion are determined in terms of $h_0$, $f_0$ and $\phi_0$. Note that the value one chooses for $f_0$ fixes the normalization of time. In the intermediate region $r_2> r> r_1$ where the fluid energy density, pressure and charge density are all non-zero,  one has to solve  the equations
\eqref{BackEq1}, \eqref{BackEq3}, \eqref{BackEq4} and
\eqref{BackEq5}. In the outer region $r\leq r_1$, the space-time is given, once again, by the solution to the equations \eqref{EintBackReacted2}-\eqref{PhiEOSBackReacted2}. The outer-region solution is asymptotically  AdS$_4$, with the leading expansion of the scalar field given by \eqref{AsymptoticPhi}. Note that in the backreacted case, the temperature of dual field theory is given by $(4\pi c)^{-1}|df(r_0)/dr\sqrt{1/f(r_0)g(r_0)}|$, where $c$ is given by $\sqrt{r^4f(r)g(r)}|_{r\to 0}$. Also,  the chemical potential of the dual theory is given by $h(r\to0)/c$.

\begin{figure}
\centering
 \hskip-0.15in \includegraphics[width=77mm]{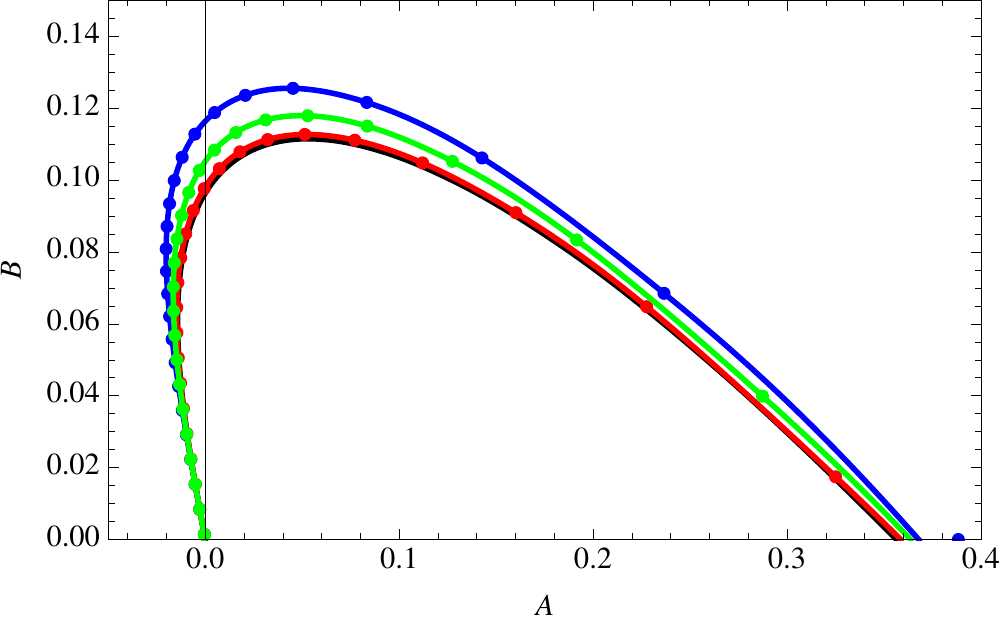}
\vskip -0.1in
\caption{\label{BoverAFTB} \footnotesize{Plots of $B$ as function of $A$. The black curve corresponds to the case without the backreaction of the scalar field, while the blue, green and red curves correspond to the scalar backreaction with $\lambda=10, 20$ and $100$ respectively. The mass square of the scalar is  $m^2L^2=-2.2$ for all the curves. For the non-backreacted case, $T/\mu \simeq0.00045$ with $m_{\rm f}=0.7$ and $\b=10$. } }
\end{figure}

The condensed phase (in the standard quantization) is  once again
characterized by $A=0$ and $B\neq0$. The plots in Figure
\ref{BoverAFTB} show the behavior of $B$ versus $A$ for different
values of the coupling $\lambda$. As the plots show, the condensed
phase persists even as $\lambda$ decreases, where the backreaction of
the neutral scalar field becomes more important. As a test of our
numerics for $\lambda\to\infty$, the curves produced with the scalar backreaction present approach the curve obtained for the case without the scalar backreaction.  We have checked that even if the scalar backreaction is included, the finite-temperature phase transition between the condensed and uncondensed phases is still second-order with mean-field exponents.

\section{Discussion}\label{SectionFive}

The Hertz-Millis \cite{hertz,millis} theory is an attempt to describe quantum criticality
in a metallic system in the presence of some type of magnetic order described by a neutral order parameter.
In obtaining the low-energy theory in terms of the bosonic neutral
order parameter, the fermions, which are gapless, are integrated out. This procedure has no guarantee of working since the fermions belong to the low-energy
sector.  In $d=2+1$, this procedure is fatal as an infinite \cite{ac,ssl,ms} number of
marginal operators are generated by the integration procedure.
Holography obviates integrating out the fermions because they can
be included non-perturbatively in the background.   While the geometry remains AdS$_4$ asymptotically, the fermions deform the space-time in the interior giving rise to an IR Lifshitz fixed-point in which the
dynamical exponent is inherently finite.  What we have shown is that
the condensation of the  neutral scalar operator in the boundary theory dual to the electron star background is controlled by the IR Lifshitz fixed-point whenever the
mass square of the neutral scalar field in the bulk violates the
BF bound.  Figure \ref{Phase} shows explicitly that the boundary theory undergoes a quantum phase transition simply by tuning the mass square of the bulk scalar field. It is the
condensation of the scalar that gives rise to the new IR scale, see equation \eqref{BKTCondensate},  that depends exponentially on the distance from the critical point.
Consequently, the underlying $T=0$ transition is of the BKT form.  At finite temperature, we find that the phase transition is
second-order and is described by mean-field exponents.  Backreaction of the condensed scalar on the geometry leads to a change in the dynamical exponent across the phase transition.  This is a particularly attractive feature of this theory as there are a variety of systems \cite{pines,fradkin} in which the condensation of a neutral field changes the dynamical exponent across the phase transition.

There are a number of immediate applications of our work.
First, our model can be used to described antiferromagnetic phases in
condensed matter systems with finite dynamical critical exponents. To
do so, we would simply follow the analysis discussed in \cite{Iqbal:2010eh} by embedding the neutral scalar field into a triplet charged under  an SU(2) gauge symmetry in the bulk; see also the discussion in \cite{Faulkner:2010tq}. This bulk SU(2) symmetry are to model the SU(2) spin symmetry in the dual boundary theory.   The antiferromagnetic ordering then  corresponds to the spontaneous breaking of the SU(2) spin symmetry to U(1) in the low energy limit. To holographically model the transition to the antiferromagnetic phase, where there is a staggered spin order parameter with zero spin density,  we introduce in the bulk an SU(2) gauge field $A^a_{M}$ along with a triplet $\phi^a$ charged under the SU(2) gauge group. Here,  $a=1,2,3$.  (Note that $\phi^a$ is neutral under the U(1) gauge group.) $A^a_{M}$ is dual to the
spin density in the boundary theory which will be set to zero, while $\phi^a$ is dual to the staggered order parameter. By embedding the neutral scalar field $\phi$ into $\phi^a$, we can explicitly break the SU(2) spin symmetry when the mass of the scalar field
falls below the BF bound of the far-interior Lifshitz region.  Further, by studying perturbations around the symmetry breaking solution, one
can find that there are two gapless Goldstone mode with linear
dispersion relations. The argument presented in \cite{Iqbal:2010eh} is
independent of the form of metric and can be applied to electron star
solution also. 

Second, we could also model nematic order.  As mentioned in the introduction, nematic  order \cite{nematic} is an instance of a
neutral tensor condensate.  The order parameter for a nematic phase
is a traceless symmetric second-rank tensor. Indeed, one way for holographically constructing such a phase would be similar to the holographic setups studied in \cite{Chen:2010mk, Benini:2010qc, Benini:2010pr} for $d$-wave superconductivity, except that the order parameter here is neutral. Thus, one introduces a massive neutral traceless symmetric rank-two tensor field in the bulk.  A rotation in the ($x,y$)-plane can be used to set one component of the tensor field equal to zero. The equation of motion for the remaining component, after a field redefinition, will be identical to the equation of motion for a neutral scalar field. So, the analysis for  the condensation of the dual tensor operator in the boundary theory is essentially mapped to the analysis for the condensation of a neutral scalar operator that we presented in this paper. It would be interesting to see the Pomeranchuk instability \cite{Pomeranchuk, Oganesyan} in this holographic setup as the quantum critical point is approached. Of course, this would require coupling the tensor field to a fermion. One should look at our heuristic description of a holographic quadrupolar nematic phase from an effective field theory point of view, mainly because, as of now, there is no consistent action describing the dynamics of a massive spin-two field which would be causal and free of ghost in a space-time which is not Einstein; see \cite{Benini:2010pr, Buchbinder:2000fy} for more discussions.

Third, there are numerous examples in condensed matter as well as particle physics (most notably in QCD) in which several ordering tendencies compete.  In the instances in which competing order involves neutral condensates in metallic systems,  the holographic method used here can be tailored to capture the relevant physics. The $\mathbb{Z}_2$ symmetry in the case of a single neutral scalar field can easily be enlarged by including more scalar fields in the action \eqref{Scalar Action} and then choosing the interactions to be invariant under the desired symmetry. Depending on which dual operators condense, a variety of interesting symmetry breaking patterns could be potentially realized in the boundary theory. 

Finally, another extension would be the inclusion of probe fermions coupled
to the neutral scalar field. For the RN-AdS$_4$ background, fermion correlators in the dual boundary theory have been studied extensively; see, for example,  \cite{Lee:2008xf, Liu:2009dm, Cubrovic:2009ye, Faulkner:2009wj, Basu:2009qz, Denef:2009yy, Chen:2009pt, Faulkner:2009am, Gubser:2009dt, Faulkner:2010tq, Gubser:2010dm, Ammon:2010pg, Benini:2010qc, Vegh:2010fc, Edalati:2010ww, Edalati:2010ge, Guarrera:2011my}. A similar kind of analysis can be done here for the
electron star background.  In fact, the results on the electron star
background would in principle be more relevant to condensed matter
systems because the IR theory is controlled by a Lifshitz fixed-point and moreover, the entropy density of the system goes to zero at zero temperature.  While some
work along these lines has been initiated in \cite{hartnollstarspec}, probe fermions which couple
to an order parameter, either charged or neutral, have not been included.  Including such couplings  could give rise
to a non-trivial dispersion for the fermions around the Fermi surface.  One interesting effect that
could be studied in this approach is whether or not a neutral
condensate with a non-trivial momentum structure could, upon
coupling to the probe fermions, give rise to a Fermi arc structure
indicative of
the pseudogap phase of the cuprates. This is particularly appealing
since most models \cite{norman} of the pseudogap involve some type of neutral order parameter. Recent holographic
constructions \cite{Benini:2010qc, Vegh:2010fc} of Fermi arcs are not particularly relevant to the
pseudogap phase of the cuprates because in these construction the existence of the Fermi arcs are due to the coupling between the probe fermion and a charged superconducting  order parameter.

\section*{Acknowledgments}
We would like to thank E. Fradkin and R. Leigh for many useful discussions, and S. Hartnoll and N. Iqbal for correspondence.  M.E., K. W.L. and P.W.P.  acknowledge financial support from the NSF DMR-0940992 and the Center for Emergent Superconductivity, a DOE Energy Frontier Research Center, Award Number DE- AC0298CH1088.

\end{document}